\documentclass[fleqn,usenatbib,useAMS]{mnras}
\usepackage[english]{babel}
\usepackage[utf8]{inputenc}
\usepackage[T1]{fontenc}
\usepackage{aecompl}
\usepackage{wrapfig}
\usepackage{amssymb}
\usepackage{graphics}
\usepackage{graphicx}
\usepackage{wrapfig}
\usepackage{epsfig}
\usepackage{amsmath}
\usepackage{bigints}
\usepackage{natbib}
\usepackage{capt-of}
\usepackage{multicol}
\usepackage{bm}
\usepackage{pdflscape}
\usepackage{physics}

\usepackage{ae,aecompl}
\usepackage{newtxtext,newtxmath}

\pdfoutput=1

\DeclareMathOperator\arctanh{tanh^{-1}}

\title[On the mean profiles of radio pulsars -- II]
{On the mean profiles of radio pulsars -- II: Reconstruction of complex pulsar light-curves and other new propagation effects}
\author[H. L. Hakobyan, V. S. Beskin and A. A. Philippov]{
H. L. Hakobyan$^{1, 2}$\thanks{Contact e-mail: \href{mailto:hakobyan@astro.princeton.edu}{hakobyan@astro.princeton.edu}},
V. S. Beskin$^{2, 3}$\thanks{Contact e-mail: \href{mailto:beskin@lpi.ru}{beskin@lpi.ru}}, 
and A. A. Philippov$^1$\thanks{Contact e-mail: \href{mailto:sashaph@princeton.edu}{sashaph@princeton.edu}} \\
$^{1}$Department of Astrophysical Sciences, Peyton Hall, Princeton University, Princeton, NJ 08544, USA\\
$^{2}$Moscow Institute of Physics and Technology, Dolgoprudny, Institutsky per., 9, Moscow Region, 141700, Russia\\
$^{3}$P.N.Lebedev Physical Institute, Leninsky prosp., 53, Moscow, 119991, Russia}

\date{Accepted 2017 April 26. Received 2017 April 05; in original form 2017 January 09}
\pubyear{2016}

\begin{document}
\label{firstpage}
\pagerange{\pageref{firstpage}--\pageref{lastpage}}
\maketitle

\begin{abstract}
Our previous paper outlined the general aspects of the theory of radio light curve and polarization formation for pulsars. We predicted the one-to-one correspondence between the tilt of the linear polarization position angle of the and the circular polarization. However, some of the radio pulsars indicate a clear deviation from that correlation. In this paper we apply the theory of the radio wave propagation in the pulsar magnetosphere for the analysis of individual effects leading to these deviations. We show that within our theory the circular polarization of a given mode can switch its sign, without the need to introduce a new radiation mode or other effects. Moreover, we show that the generation of different emission modes on different altitudes can explain pulsars, that presumably have the X-O-X light-curve pattern, different from what we predict. General properties of radio emission within our propagation theory are also discussed. In particular, we calculate the intensity patterns for different radiation altitudes and present light curves for different observer viewing angles. In this context we also study the light curves and polarization profiles for pulsars with interpulses. Further, we explain the characteristic width of the position angle curves by introducing the concept of a wide emitting region. Another important feature of radio polarization profiles is the shift of the position angle from the center, which in some cases demonstrates a weak dependence on the observation frequency. Here we demonstrate that propagation effects do not necessarily imply a significant frequency-dependent change of the position angle curve.
\end{abstract}


\section{Introduction}

During almost fifty years of study from the very beginning in 1967, when radio pulsars were first observed~\citep{PSRdiscovery}, the major understanding in neutron stars' magnetosphere structure and in the origin of their activity was achieved~\citep{L&GS, L&K}. However, some key questions including the mechanism of the coherent radio emission generation still remain unexplained. The mass $M$, the period $P$, and the breaking factor of the pulsar $\dot P$ can be determined directly with a good accuracy, but, on the other hand, such important parameter as the inclination angle $\alpha$  between magnetic and rotational axes can be found only using the so-called rotating vector model of the position angle swing along the mean radio profile ~\citep{L&M, T&M, Rookyard15}. However, this measurement is usually non-reliable, as the main assumption of RVM, e.g. polarization of radio emission is formed in the generation region at distance $\sim 10\text{-}30R$ from the stellar surface (with $R$ here and further being the radius of the neutron star), is not justified. This is mainly because radio emission interacts with plasma created in magnetospheric discharges, and radiation polarization characteristics start to deviate from the simple prediction of RVM. In order to make correct theoretical predictions of the radio polarization all propagation effects, e.g. magnetospheric plasma birefringence \citep{B&A, BGI88, P&L1, P&L2}, cyclotron absorption ~\citep{mikhailovskii82, abs1, luomelrose2004}, and limiting polarization ~\citep{C&R, lim1, P&L2, lim3, andrianovbeskin2010, wanglaihan2010}, should be accurately taken into account.

This is the second paper dedicated to the study of polarization characteristics based on the quantitative theory of the radio waves propagation in the pulsar magnetosphere. In Paper I~\citep{beskinphilippov2012} the theoretical aspects of the polarization formation based on~\citet{kravtsovorlov90} approach were studied, and the numerical simulation method was proposed. It allowed us to describe the general properties of mean profiles such as the position angle of the linear polarization $p.a.$ and the circular polarization for the realistic structure of the magnetic field in the pulsar magnetosphere. We confirmed the main theoretical prediction found by~\citet{andrianovbeskin2010}, i.e., the correlation of signs of the circular polarization, $V$, and derivative of the position angle with respect to pulsar phase, ${\rm d} {p.a.}/{\rm d}\phi$ for both emission modes. In most cases it gave us the possibility to recognize the orthogonal mode, ordinary or extraordinary, playing the main role in the formation of the mean profile.

On the other hand, there are some pulsars for which observations were in disagreement with our predictions. The detailed statistical analysis of polarization characteristics that support the predicted O-X-O light-curve model will be presented in Paper III (Jaroenjittichai et al. in preparation), while in a current paper we focus on more detailed analysis of the wave propagation in the pulsar magnetosphere. In Sect.~\ref{sec:propeff} and~\ref{sec:cycabs} we briefly discuss the propagation model and other theoretical assumptions used in our simulations. We compare our model with the broadly used geometric models, namely, the hollow cone model and the rotating vector model with the aberration/retardation effects, to emphasize the features that are different from that simplified model.

In consequent sections we discuss the results obtained using our technique. Sect.~\ref{sec:anomal} is dedicated to the above mentioned deviations from our predictions. In Sect.~\ref{sec:vsign} we explain the switch of the circular polarization sign of a single mode, that was not predicted previously, but is clearly visible for some pulsars. The predicted O-X-O mode sequence is seen to be broken in some of the pulsars' profiles. We show in Sect.~\ref{sec:anomalstokes} that if the two modes are being generated at different heights, one can indeed explain this anomalous behavior. In Sect.~\ref{sec:chump} we briefly discuss the possibility to explain the central hump in the position angle curves of some of the pulsars. It is demonstrated, that there is no need to introduce a complicated altitude profile of the radiation.

In Sect.~\ref{sec:genprop} we focus on some general properties of the formation of light curves and polarization profiles, namely for ordinary pulsars (Sect.~\ref{sec:dirpatt}) and for the pulsars with interpulses (Sect.~\ref{sec:interpulse}). For both cases we show the intensity pattern in the picture plane and the Stokes parameter map at a given altitude and explain how different profiles can be formed in this context. In Sect.~\ref{sec:radtofreq} we explain the width of the position angle curves. Finally, we discuss the shift of the position angle on different frequencies with the propagation effects taken into account in Sect.~\ref{sec:posangshift}.

\section{Propagation theory}
\label{sec:propth}

In this section we remember general assumptions about the radiation generation and propagation effects that we use for our calculations. We also discuss some important results obtained in Paper I.

\subsection{Hollow cone model}

For a long time it was known, that there are two orthogonal modes propagating in pulsars' magnetosphere: the extraordinary X-mode and the ordinary O-mode~\citep{T&S}. While the X-mode propagates along the straight line without any refraction, the O-mode is being deflected from the magnetic axis~\citep{B&A, BGI88, P&L1, P&L2}. This led to the idea of the modification of the hollow-cone model for the directivity pattern generation, where there is an inner cone --- straightly propagating X-mode, and the outer one is the O-mode that is deflected from the magnetic axis~\citep{BGI93}. Radiation in the central region of the cone is suppressed due to large curvature radius of the magnetic field lines, as well as outside the edges of the polar cap region, where there are no open field lines. Various pulsar profiles correspond to different intersections of the line of sight and the directivity pattern from the hollow-cone model. Note, that in Sect.~\ref{sec:dirpatt} we show, that in fact the directivity pattern can be more complex depending on various plasma parameters. \par

Remember that the 'hollow cone' model assumes the magnetic field to be dipolar with the radiation propagating along a straight line. The polarization characteristics themselves are formed exactly in the same region, where the radiation is generated, i.e., deep near the stellar surface. This assumptions allow us to analytically calculate the so-called Rotating Vector Model (RVM) curve for the $p.a.$ plot along the rotation phase $\phi$~\citep{radhakrishnan69},
\begin{equation}\label{eq:rvm}
    p.a.=\arctan{\left(\frac{\sin{\alpha}\sin{\phi}}{\sin{\alpha}\cos{\zeta}\cos{\phi}-\sin{\zeta}\cos{\alpha}}\right)},
\end{equation}
where $\zeta$ is the angle between the rotation axis and the line of sight. Equation (\ref{eq:rvm}) can be obtained considering the linear polarization that rigidly follows the magnetic field direction. Note, that in this paper, dissimilar to Paper I, the impact angle $\beta$ is chosen as $\beta = \zeta - \alpha$, i.e., negative $\beta$ corresponds to the line of sight closer to the rotation axis, than the magnetic moment. On the other hand, similar to Paper I, the sign of the $p.a.$ is conventionally (astronomically) chosen, see, e.g.,~\citet{evweis2001}. In this case the $\Theta_1$ variable from~\citet{kravtsovorlov90} equations (see below) corresponds exactly to the position angle of the linear polarization.

\subsection{Propagation effects}
\label{sec:propeff}

While neglecting the propagation effects can work for some pulsars, most of them, however, appear to poorly correspond to this simplified approach. First of all, the $p.a.$ curves of some profiles appear to be shifted from the center of the profile (clearly breaking the RVM-curve) and some of them, e.g., PSR~J1022+1001, expose anomalous humps in the center~\citep{gouldlyne98}. This problem is usually solved by considering the so-called aberration/retardation effects (later A/R) and by the assumption that the radiation is generated at some particular altitude~\citep{blaskiewicz91, mitra2004, dyks2008, mitragupta2009}. The A/R effect allows to determine the shift of the $p.a.$ curve as $\Delta\phi\approx 4r_{\rm em}\Omega/c$ and hence deduce the radiation origin height $r_{\rm em}$. The general agreement from this simplified technique, which is in a good consistency with geometric conclusions, is that the radiation originates in the deep regions near $(10\text{-}100)$ stellar radius. However, it is clear, that to address to this problem self-consistently, one must take into account  propagation effects in the neutron star magnetosphere. 

On the other hand, some profiles expose a nontrivial circular polarization and even a polarization sign reversal, not only in the core emission, but in the conal part as well (see~\citealt{hanmanxuqiao1998}). The early papers~\citep{C&R, HL&K} proposed a possible explanation of the circular polarization by assuming a wave propagation at a nonzero angle to the finite magnetic field line. However, for the radiation formation in deep regions, where the magnetic field is high enough, these explanations failed to work. 
A more accurate consideration of the circular polarization formation in the limiting polarization region in ultrarelativistic highly-magnetized magnetosphere by~\citet{P&L3} provides a possible explanation, however circular polarization is not yet explained quantitatively without sticking to a particular radiation mechanism (see, e.g., \citealt{wangwanghan2012}).\par

As it was already stressed, the importance of the propagation effects was shown by~\citet{B&A, BGI88, P&L1, P&L2}. First, the refraction of the O-mode takes place in the region $r < r_\mathrm{O}$, where
\begin{equation}\label{eq:rO}
    r_\mathrm{O}\sim10^2R\cdot\lambda_4^{1/3}\gamma_{100}^{1/3}B_{12}^{1/3}\nu_\mathrm{GHz}^{-2/3}P^{-1/5}.
\end{equation}
Here and below $R$ is the stellar radius, $\lambda$ is the multiplicity parameter
\begin{equation}\label{eq:lambda}
    \lambda = \frac{n_{\rm e}}{n_{\rm GJ}},
\end{equation}
i.e., the electron-positron number density normalized to Goldreich-Julian one ($\lambda_4 = \lambda/10^4$), $\gamma_{100}$ is the characteristic Lorentz-factor of secondary plasma normalized by $10^2$, $B_{12}$ is the polar cap magnetic field $B_{0}$ in $10^{12}~\text{G}$, $\nu_\mathrm{GHz}$ is the frequency in GHz and P is the period of rotation in seconds.\par

On the other hand, as the number density $n_{\rm e}$ quickly decreases far from the star surface, the ray transits from the region of the dense plasma where the linear polarization follows the external magnetic field, to the region of rarefied plasma where the external magnetic field cannot affect the polarization of a ray. As a result, the polarization freezes at some distance $r_{\rm esc}$ (so-called limiting polarization, see \citealt{zhelezniakov77}). For ordinary pulsars one can obtain~\citep{C&R, andrianovbeskin2010}
\begin{equation}\label{eq:resc}
    r_{\rm esc}\sim 10^3R\cdot\lambda_4^{2/5}\gamma_{100}^{-6/5}B_{12}^{2/5}\nu_{\rm GHz}^{-2/5}P^{-1/5}.
\end{equation}
As we see, for ordinary radio pulsars the escape region is located well inside the light cylinder $R_{\rm L} = c/\Omega \approx 10^{4} R$, but much higher than the radiation domain. Thus, one should consider the evolution of polarization characteristics from the generation region $r_{\rm em}$ up to the altitude $r = r_{\rm esc}$ at which the polarization freezes.\par

In Paper I the numerical approach with the method of~\citet{kravtsovorlov90} equation was proposed that describes the evolution of polarization characteristics along the line of sight on complex angle $\Theta=\Theta_1+i\Theta_2$, with $\Theta_1$, in agreement with $p.a.$ determination (\ref{eq:rvm}), being the $p.a.$ and \mbox{$\Theta_2=1/2\arctanh{V/I}$,} where $V/I$ is the relative level of circular polarization. For small $\Theta_2 \ll 1$ one can approximate the level of circular polarization as 
\begin{equation}
    \frac{V}{I}\propto\frac{\mathrm{d}\left(\beta_B+\delta\right)/\mathrm{d}l}{\cos\left[2(p.a.-\beta_B-\delta)\right]},
\label{eq:VI}
\end{equation}
where the derivative is taken near the $r_{\rm esc}$. Here the angle $\beta_B$ corresponds to orientation of the external magnetic field in the picture plane and the additional phase $\delta$ appears due to the external electric field resulting in electric drift motion of particles in the pulsar magnetosphere 
\begin{equation}
\tan{\delta} =  -\frac{\cos\theta \, U_{y}/c}{\sin\theta - U_{x}/c}.
\label{delta}
\end{equation}
Here $U_{x}$ and $U_{y}$ are two components of the $\bmath{E} \times \bmath{B}$ drift velocity, and $\theta$ is the angle between wave vector $\bmath{k}$ and external magnetic field $\bmath{B}$. It is the phase $\delta$ that is responsible for the aberration in this approach. Note, that if the propagation effects are neglected, we have the position angle following the direction of the magnetic field, i.e., $p.a.=\beta_B$ for ordinary and $p.a.\approx \beta_B + \pi/2$ for extraordinary mode. According to (\ref{eq:VI}), it gives opposite signs for Stokes parameter $V$ for two orthogonal modes. \par

\subsection{Cyclotron absorption}
\label{sec:cycabs}

\begin{figure}
\centering
\includegraphics[width=\columnwidth]{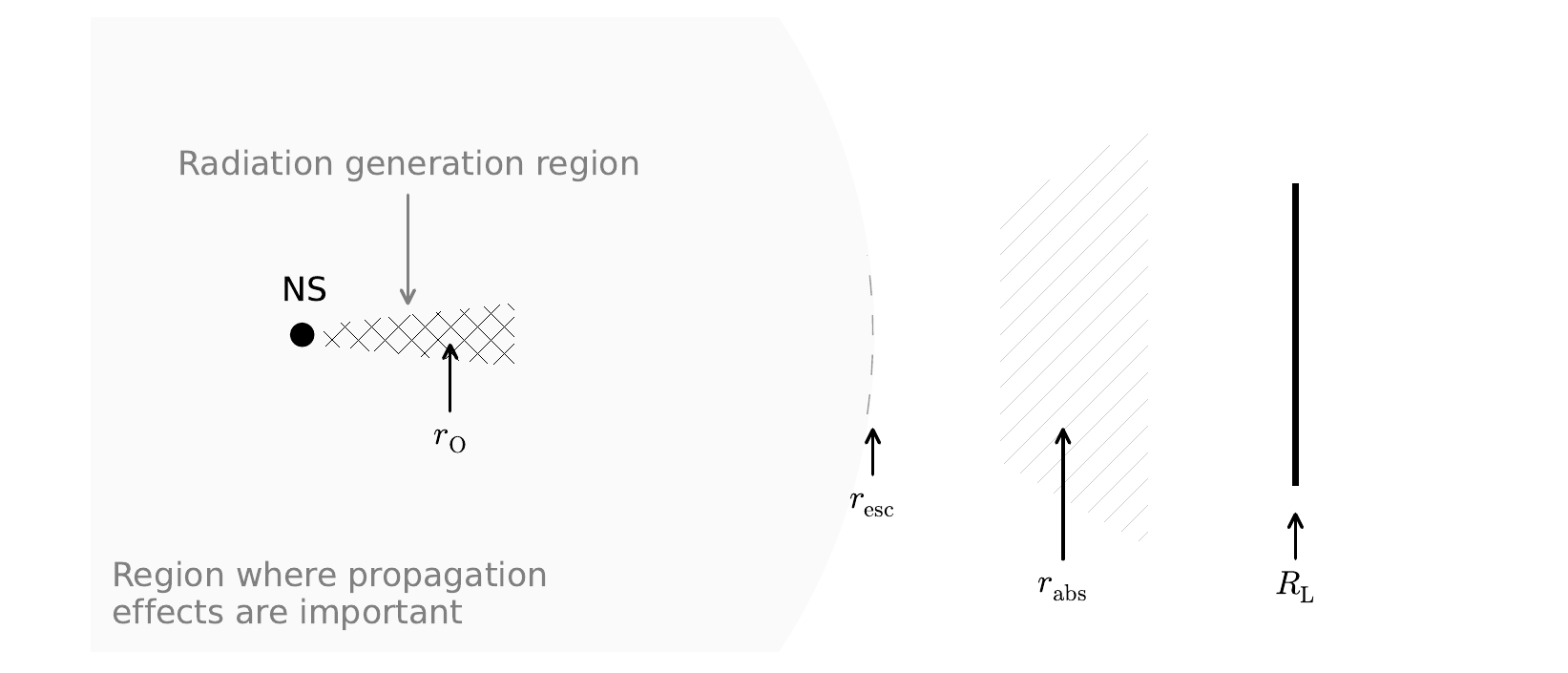}
\caption{Schematic illustration of the wave propagation in the magnetosphere, emphasizing where the particular propagation effects take place. $r_{\rm O}\sim10R$ is the height, above which the deflection of the O-mode can be neglected, the polarization remains constant above $r_{\rm esc}\sim 0.1 R_{\rm L}$ and the cyclotron resonance takes place at $r_{\rm abs}\sim R_{\rm L}$.}
\label{fig:dists}
\end{figure}

Cyclotron absorption takes place in the region where the resonance condition $\omega_{\rm B} = \gamma \tilde{\omega}$ holds. Here $\tilde{\omega}$ is the shifted frequency, i.e., $\tilde{\omega}=\omega-\bmath{k} \cdot \bmath{v}$. The distance from the stellar surface at which the resonance takes place can be found as~\citep{mikhailovskii82}
\begin{equation}
\label{eq:rabs}
    r_{\rm abs} \approx 1.8\times 10^3R\cdot\nu_{\rm GHz}^{-1/3}\gamma_{100}^{-1/3}B_{12}^{1/3}\theta_{\rm abs}^{-2/3}.
\end{equation}
Here $\theta_{\rm abs}$ is the angle between the propagation line and local magnetic field. As a result, the intensity of a ray at large distance $I_{\infty}$ can be expressed through its initial intensity $I_0$ by clear connection $I_{\infty}=I_0 e^{-\tau}$ with the optical depth 
\begin{equation}
\label{eq:depth1}
    \tau=\frac{2\omega}{c}\bigintsss_{r_{\rm em}}^{> r_{\rm abs}} \mathrm{Im}\left[n\right]\mathrm{d}l.
\end{equation}
Here $r_{\rm em}$ is the generation height and $n$ is the refractive index, found by averaging the dielectric tensor over the plasma distribution function. For a given choice of the energy distribution function $F(\gamma)$ we obtain~\citep{blandford76, luomelrose2004}
\begin{equation}
 \label{eq:depth}
 \tau \approx \frac{\pi\omega}{c}\bigintss_{r_{\rm em}}^{>r_{\rm abs}}\frac{\omega_{\rm p}^2}{\omega^2}F\left(\frac{|\omega_{B}|}{\tilde{\omega}}\right)\mathrm{d}l.
\end{equation}
 It is also useful to write down the approximate simple expression~\citep{mikhailovskii82}
\begin{equation}\label{eq:dsimple}
    \tau\approx\lambda(1-\cos{\theta_{\rm abs}})\frac{r_{\rm abs}}{R_{\rm L}}.
\end{equation}
As will be shown below, cyclotron absorption plays one of the main roles in formation of the mean profile of radio pulsars.\par

\begin{figure}
\centering
\includegraphics[width=\columnwidth]{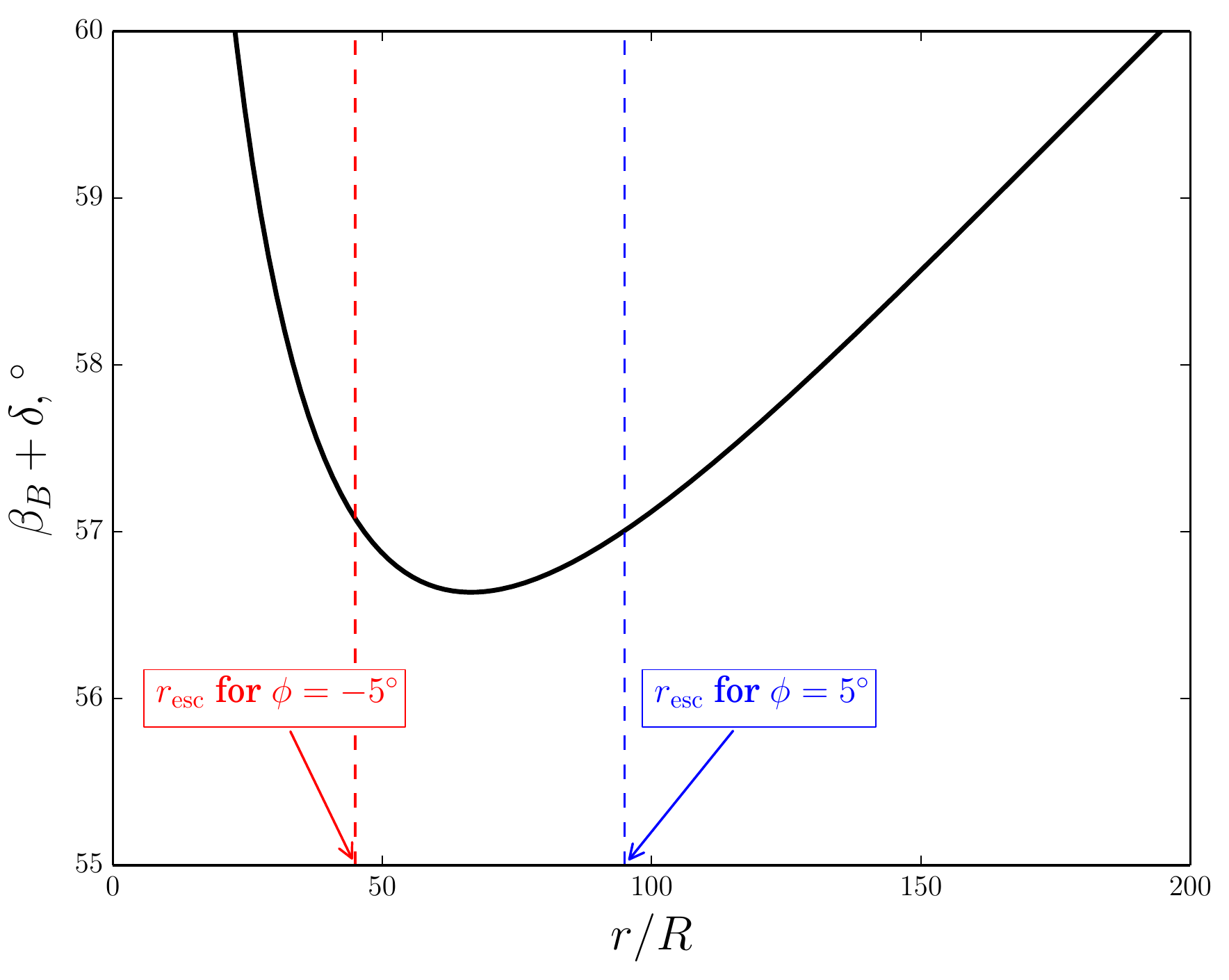}
\caption{The plot of $\beta_B+\delta$ along the propagation ray. If the polarization is formed below the extremum, the sign is governed by the derivative of $\beta_B$, and when it's formed above - the sign is determined by the derivative of $\delta$ (Paper I).}
\label{fig:betadelta}
\end{figure}

\begin{figure}
\centering
\includegraphics[width=\columnwidth]{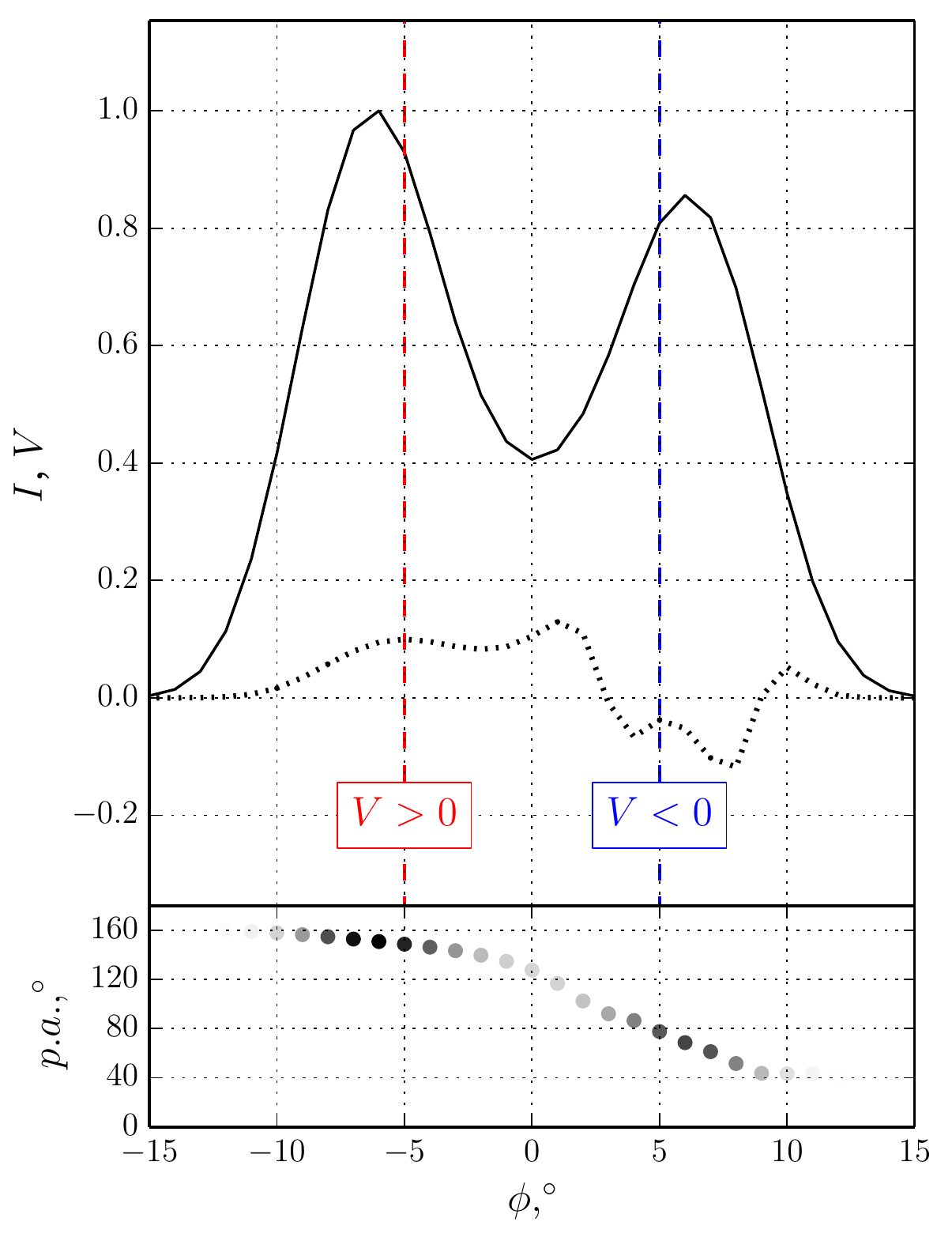}
\caption{The profile for a pulsar with high $\gamma_0$. In this case for different phases $\phi=-5\degr$ and $\phi=5\degr$ the circular polarization has different signs, due to the variation of the escape radius with respect to the extremum of $\beta_B+\delta$ (see Fig.~\ref{fig:betadelta}).}
\label{fig:vsign}
\end{figure}

To summarize, let us enumerate the main propagation effects in terms of distances from the neutron star that are to be taken into account (see Figure~\ref{fig:dists}).
\begin{enumerate}
\item The radiation will origin at some level $r =  r_{\rm em}$, which is a free parameter in our consideration. 
\item For $r<r_{\rm O}$ (\ref{eq:rO}) the refraction of O-mode takes place; for ordinary pulsars  $r_{\rm O} \sim  (20 \text{-} 50)R$. As this level depends on the frequency $\nu$, the final directivity pattern of the O-mode depends essentially on the radius-to-frequency mapping. E.g., for frequency-independent radiation radius $r_{\rm em}$ one can obtain for the frequency dependence of the mean pulse window width $w_{\rm O} \propto \nu^{-0.14}$~\citep{BGI88}. 
\item The polarization evolves until the region of limiting polarization $r\sim r_{\rm esc}$ (\ref{eq:resc}), and to reproduce it correctly, one should integrate the Kravtsov-Orlov system at least until this height.
\item For most of the pulsars the light cylinder radius $R_{\rm L}=c/\Omega$ is large enough and polarization usually forms before reaching this region, i.e., $r_{\rm esc} < R_{\rm L}$. However for millisecond pulsars or for pulsars with high plasma multiplicity $\lambda$ the limiting polarization region $r_{\rm esc}$ can be comparable or even exceed $R_{\rm L}$. In this case it is important to take into account the quasi-monopole component of the magnetic field.
\end{enumerate}

\section{New effects}
\label{sec:anomal}

\subsection{Sign switch of the circular polarization}
\label{sec:vsign}


\begin{figure*}
\centering
\includegraphics[scale=0.55]{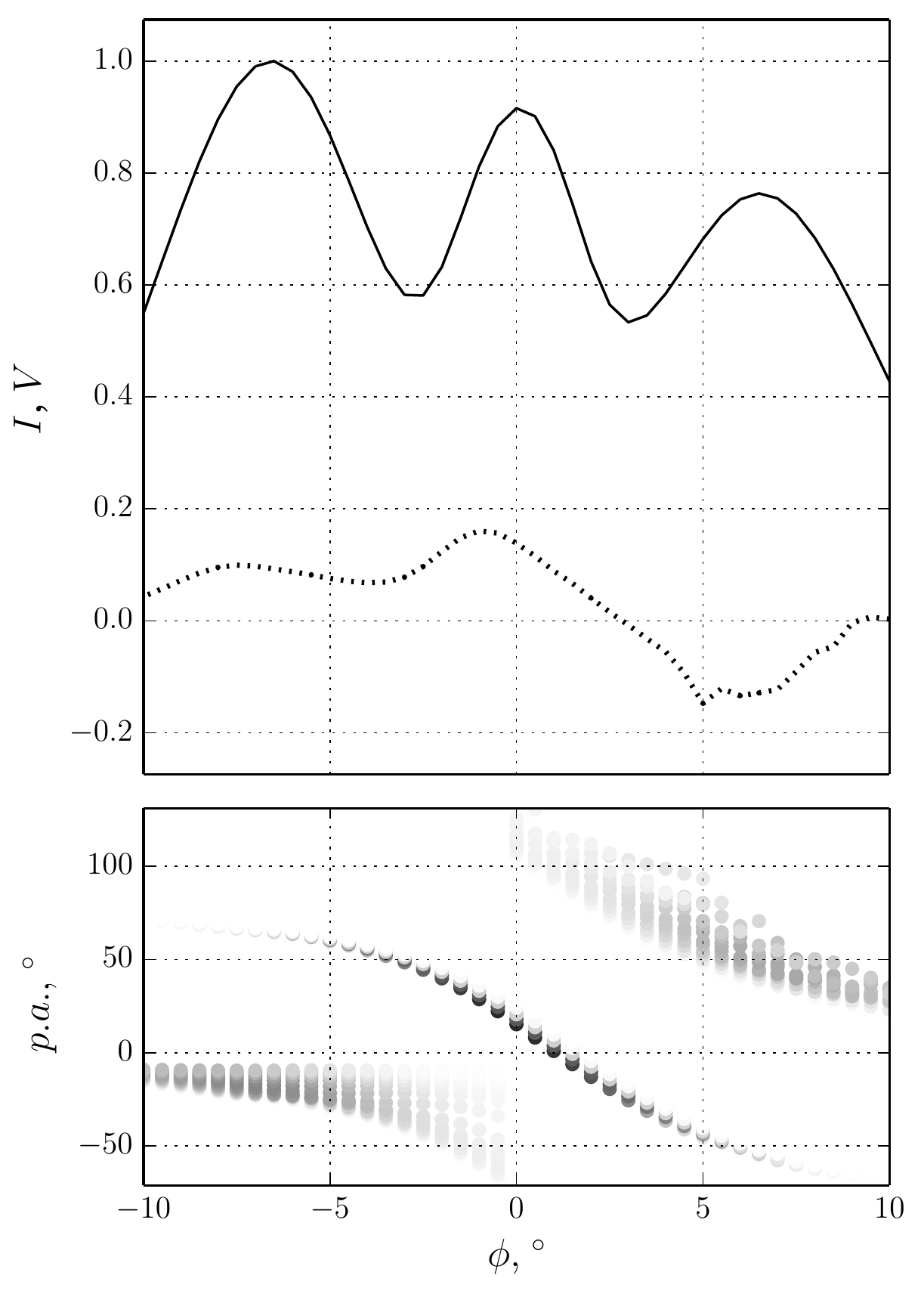}
\includegraphics[scale=0.55]{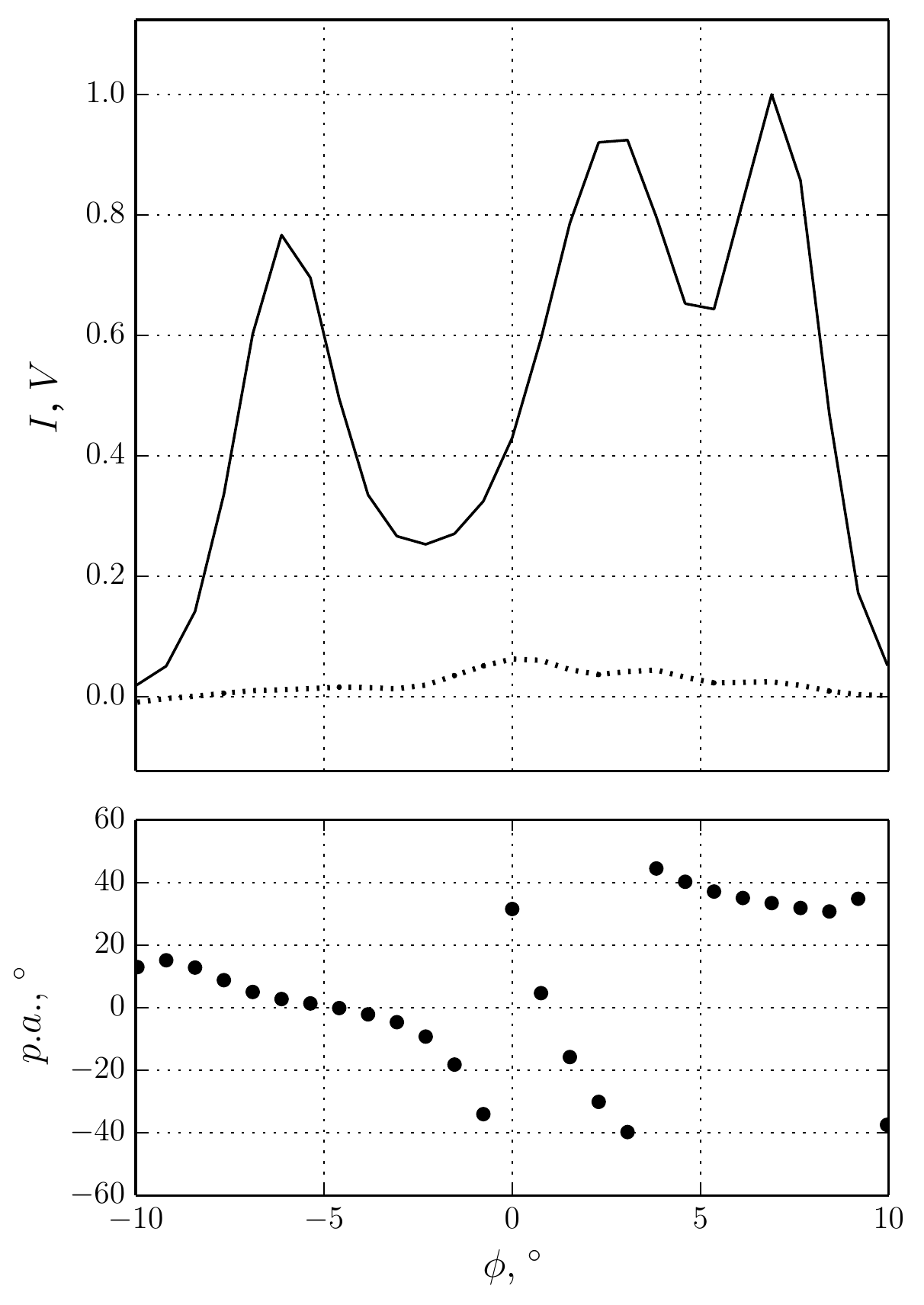}
\caption{Simulated profile for PSR~J2048-1616 at 410 MHz (left) in comparison with observational data (right) from \citet{gouldlyne98}.}
\label{fig:j2048}
\end{figure*}

\begin{figure*}
\centering
\includegraphics[scale=0.55]{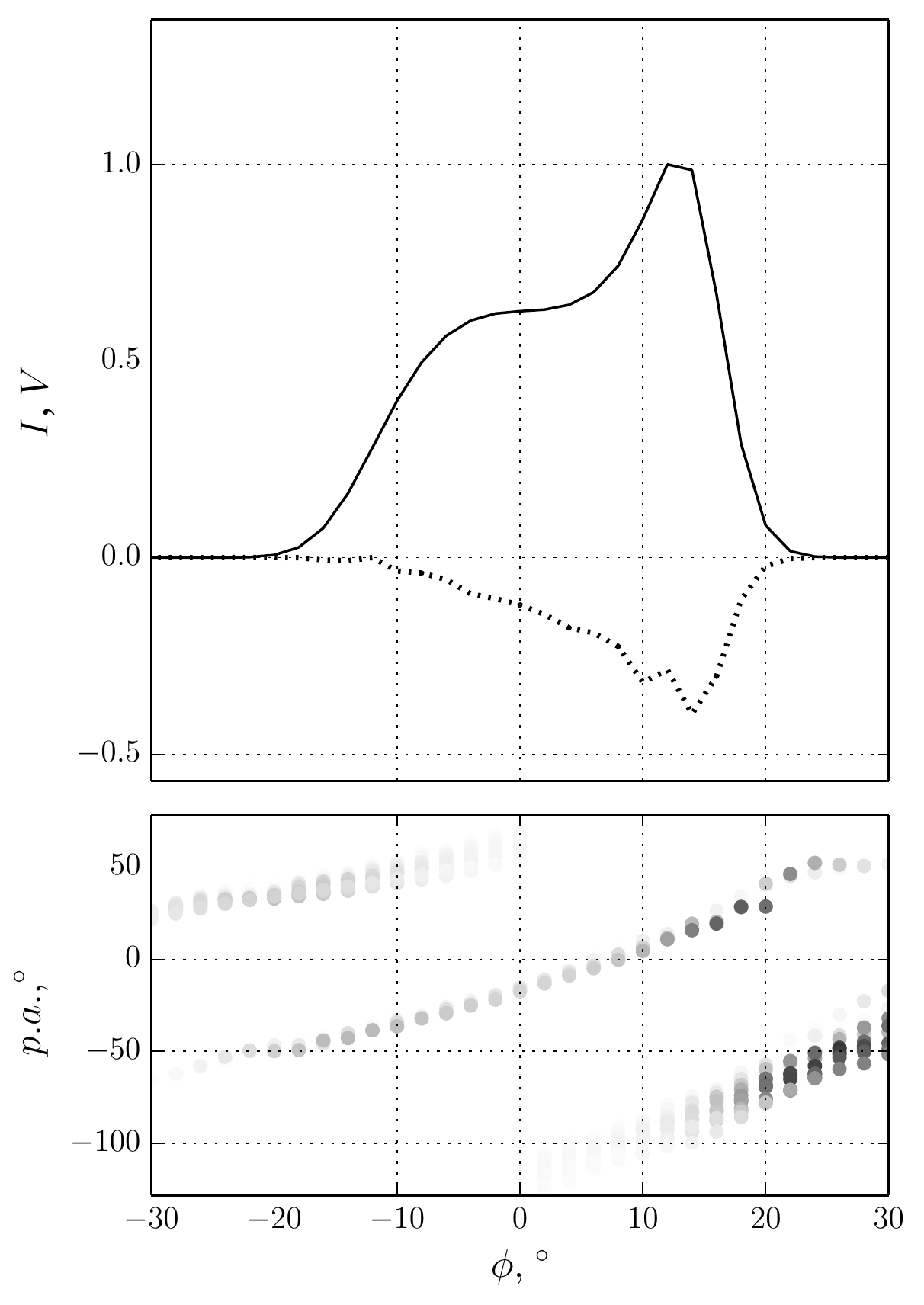}
\includegraphics[scale=0.55]{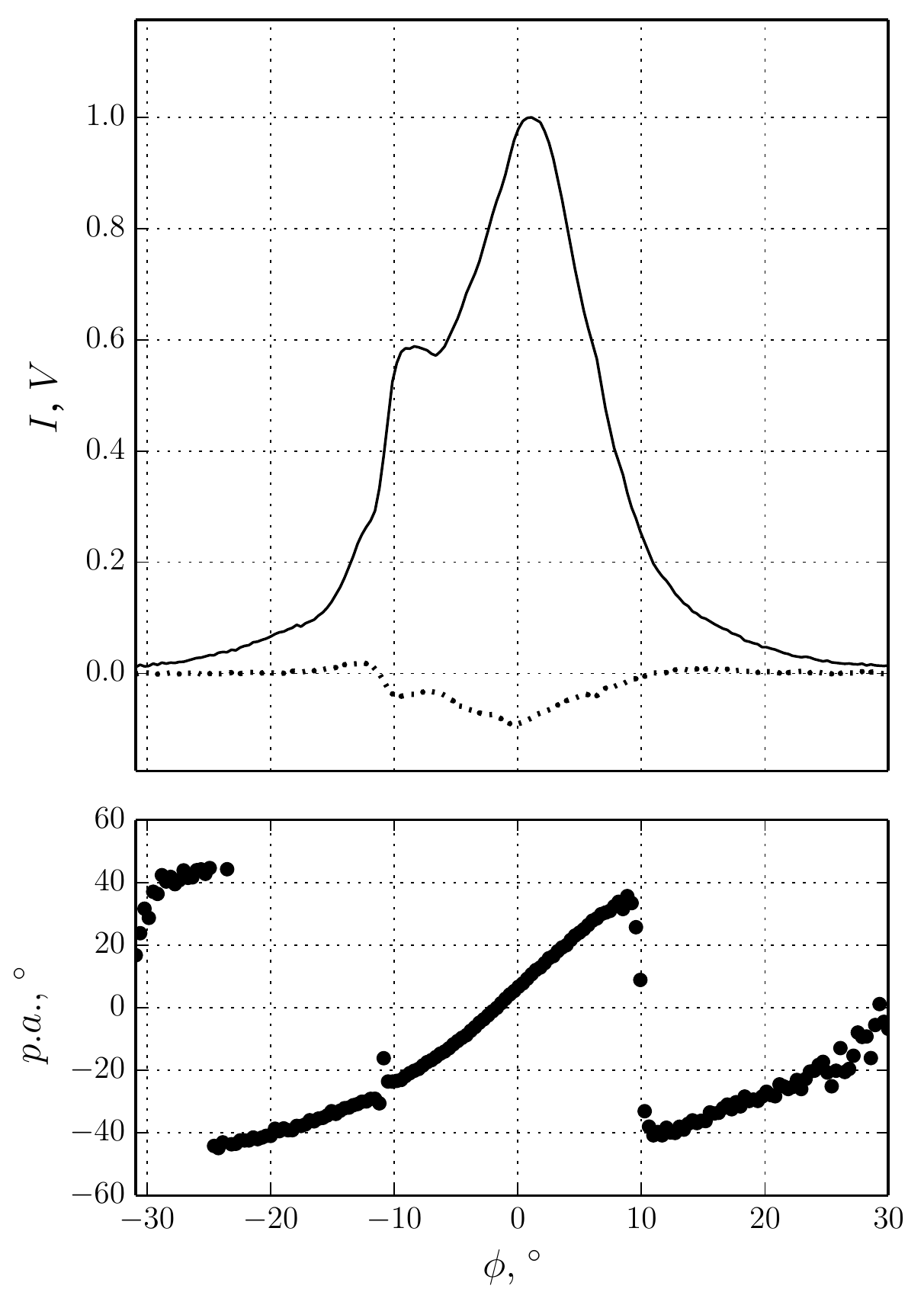}
\caption{Simulated profile for PSR~J0738-4042 at 1375 MHz (left) in comparison with observational data (right) from \citet{karjohnston2006}.}
\label{fig:j0738}
\end{figure*}

As one can see from (\ref{eq:VI}), $V\propto \mathrm{d}(\beta_B+\delta)/\mathrm{d}l$ meaning a one-to-one correspondence between the sign of the circular polarization and the derivative of $(\beta_B+\delta)$. On the other hand, as is shown in Figure~\ref{fig:betadelta}, the derivatives $\mathrm{d}\beta_B/\mathrm{d}l$ and $\mathrm{d}\delta/\mathrm{d}l$ along the ray have different signs: while at lower altitudes the first term prevails \citep{andrianovbeskin2010, wanglaihan2010}, at higher altitudes $\beta_B$ is nearly constant, and the sign is dictated by the derivative \mbox{of $\delta$.} 

In Paper I we found that in most cases the sign of the derivative $\mathrm{d}(\beta_B+\delta)/\mathrm{d}l$ is governed by $\delta$, i.e., $V\propto\mathrm{d}\delta/\mathrm{d}l$. On the other hand, as was shown by~\citet{andrianovbeskin2010, wanglaihan2010}, the sign of the derivative $\mathrm{d}\beta_B/\mathrm{d}l$ coincides with the sign of the observable derivative $\mathrm{d}p.a./\mathrm{d}\phi$. This leads to conclusion that the signs of $V$ and $\mathrm{d}p.a./\mathrm{d}\phi$ are correlated: for X-mode they are the same and opposite for O-mode. 

However, in general the sign of $V$ is sensitive to the escape radius $r_{\rm esc}$. If this radius is well above the extremum of $\beta_B+\delta$ (see Figure~\ref{fig:betadelta}), then the sign of $V$ is fixed during the whole profile. However, when the $r_{\rm esc}$ is near the extremum, polarization can be formed slightly below or slightly above this altitude, since plasma density along the ray can vary with phase. This results in different signs of $V$ for different phases.\par 

Note, that in Fig.~\ref{fig:betadelta} the polarization is being formed at different heights for phases $\phi=-5\degr$ and $\phi=5\degr$ resulting in different signs of $V$ (see Fig.~\ref{fig:vsign}). This can be the case for high Lorentz-factors of the secondary plasma, since $r_{\rm esc}$ is most sensitive to $\gamma_0$ (\ref{eq:resc}). As it will be shown in Paper III, it is this point that helps us to explain some of the exceptional profiles, for example, PSR J2048-1616.\par

\subsection{ Deviations from the predicted mode sequence}
\label{sec:anomalstokes}

If two orthogonal modes are detected in the three-component mean profile, they are presumably in the O-X-O sequence. In Paper III we critically confront this prediction with polarization data collected by~\citet{WJ2008} and~\citet{hankinsrankin2010} and demonstrate general agreement between the predictions of the theory and observations.\par

However, some pulsars demonstrate polarization profiles that poorly fit to this simplified model. Namely, at high frequencies PSR~J2048-1616 has three peaks following ( due to the sign of circular polarization) the X-O-X pattern while $p.a.$ curve shows only one orthogonal mode. On the other hand, in PSR~J0738-4042 $p.a.$ data indicates two orthogonal modes while the circular polarization $V$ does not change its sign. In Figure~\ref{fig:j2048} and Figure~\ref{fig:j0738} we show the simulated profiles and observation data for PSR~J2048-1616 and PSR~J0738-4042. In these calculations we assume that the O-mode in both cases is being generated deep enough near the stellar surface, resulting in the core component of the three peaked profile, while the X-mode is emitted further above at higher altitudes forming the edges of the pattern. In this case, for specific radiation altitudes, one would expect to have an X-O-X pattern. \par


\begin{figure}
\centering
\includegraphics[scale=0.55]{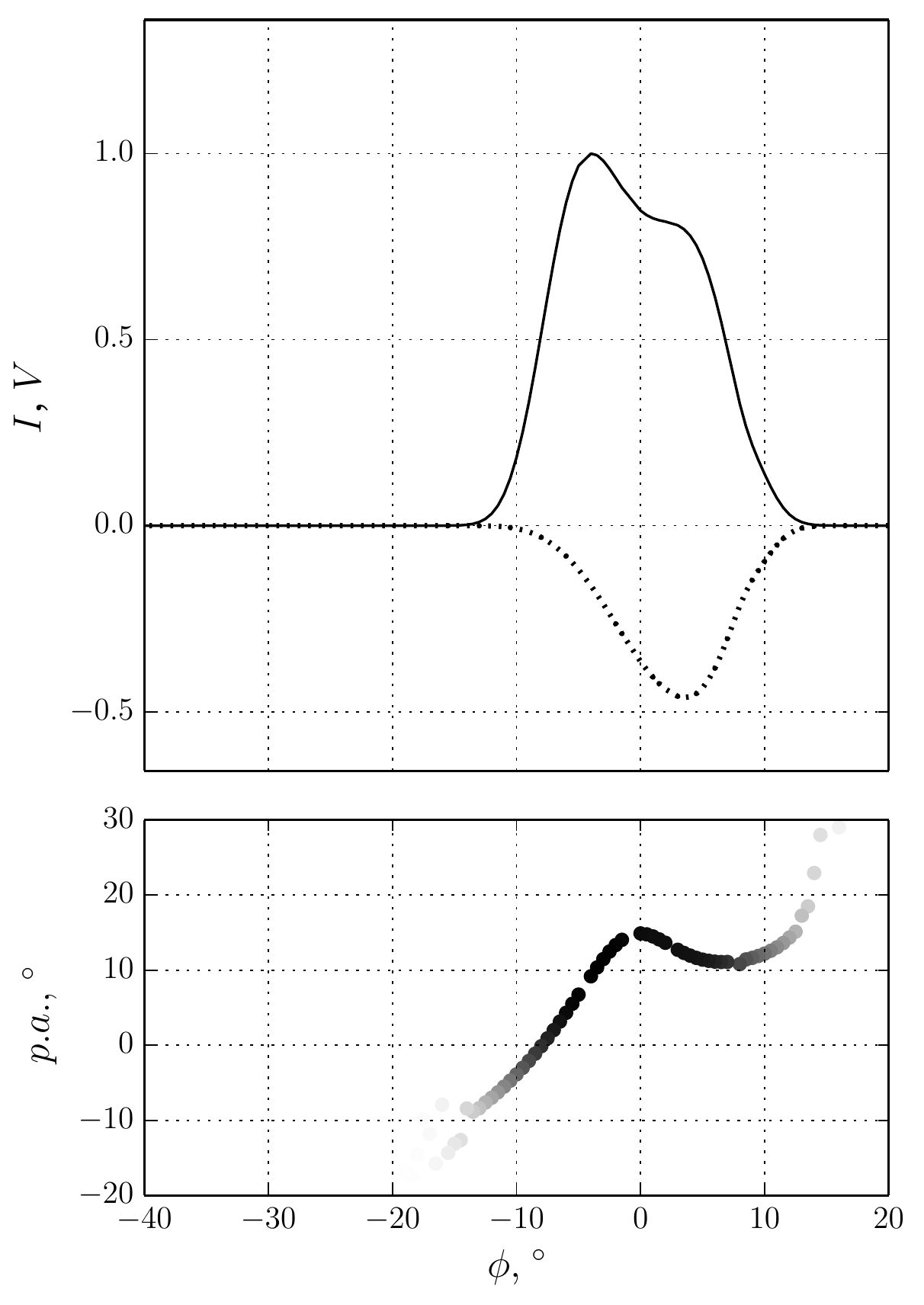}
\includegraphics[scale=0.55]{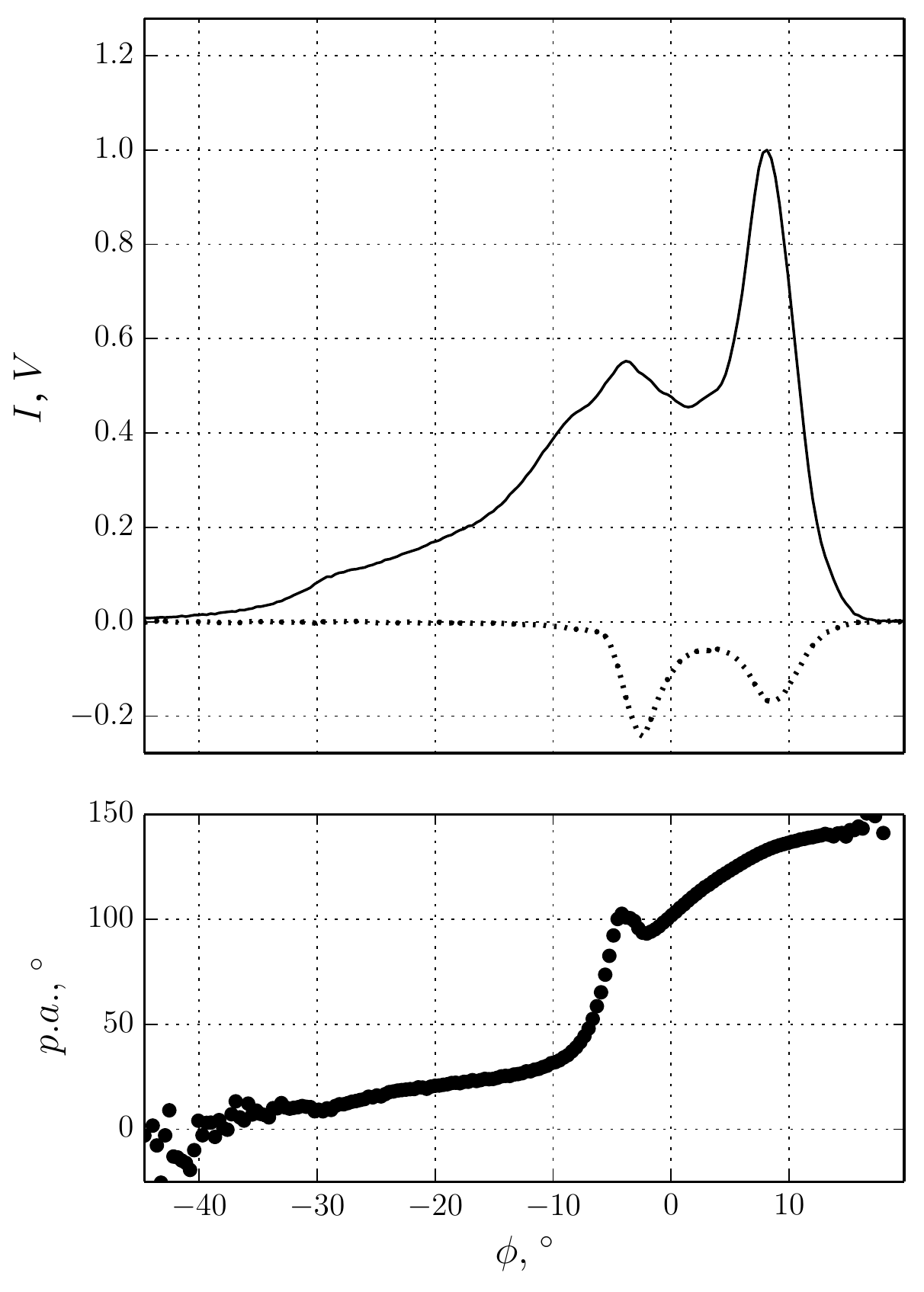}
\caption{Simulated (upper) hump in the position angle curve for the profile of PSR~J1022-1001 at 728 MHz in comparison with observational data (lower) from \citet{daihobbs2015}.}
\label{fig:j1022}
\end{figure}

\subsection{Central hump}
\label{sec:chump}

Thus, we model the radiation on two distinct widely separated regions. The altitude parameters for both pulsars are given in Table~\ref{tab:anomal}. As one can see, the anomalous polarization profiles can be qualitatively explained using this technique. In fact, the same approach can be used to explain the profile and polarization curve for another pulsar, PSR~J1146-6030, exposing similar polarization pattern. However, it is important to note, that the absolute intensity of the radiation from a given radius is an open parameter that in this case was adjusted empirically to fit the profiles.

As it was mentioned above, some two-peaked pulsars demonstrate a strange $p.a.$ behaviour at the central region (see Figure~\ref{fig:j1022}, right panel). This hump behaviour was previously discussed by~\citet{mitra2004} where the authors assumed that radiation is generated at various heights. As $\Delta p.a.\sim 4 \Omega r_{\rm em}/c$ resulting from R/A effect, they solved the inverse problem and reconstructed the complex radiation altitude profile. In this paper we show that there is no need to assume an anomalous altitude profile to explain this property.\par

Indeed, the difference of $p.a.$ curve from the standard RVM curve (\ref{eq:rvm}) is as strong as the density of secondary plasma along the ray. Thus, in the regions, where the plasma density is suppressed (i.e., in the central region of the 'hollow cone') our curve will tend to be closer to RVM one, resulting the hump in the center of the $p.a.$ curve. This phenomena can as well be observed in some two-peaked profiles near the central region \citep{WJ2008}. However, due to suppression of radiation in that region, it is hard to detect the $p.a.$ value there. \par

\begin{table}
\caption{Radiation region for PSR~J2048-1616 and PSR~J0738-4042.}
\label{tab:anomal}
 \begin{tabular}{ccc}
 \hline
 Mode & $r_{\rm em}/R$ & $\Delta r_{\rm em}/R$ \\
 \hline
 \hline
X-mode & 100 & 40 \\ 
 \hline
 O-mode & 10 & 5 \\ [1ex] 
 \hline
\end{tabular}
\end{table}


\begin{figure*}
\centering
\includegraphics[scale=0.35]{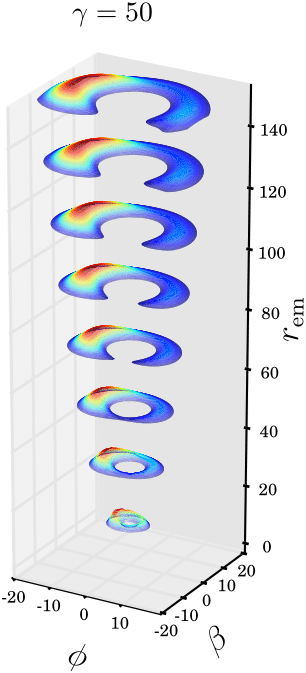}
\includegraphics[scale=0.35]{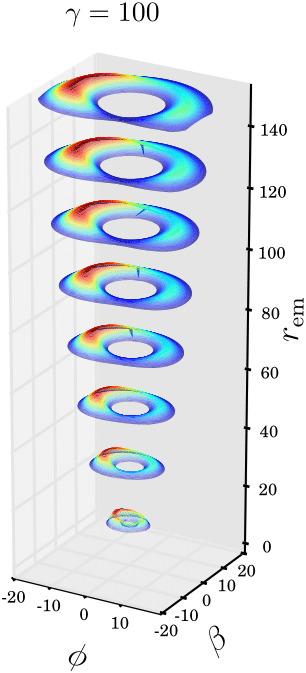}
\includegraphics[scale=0.35]{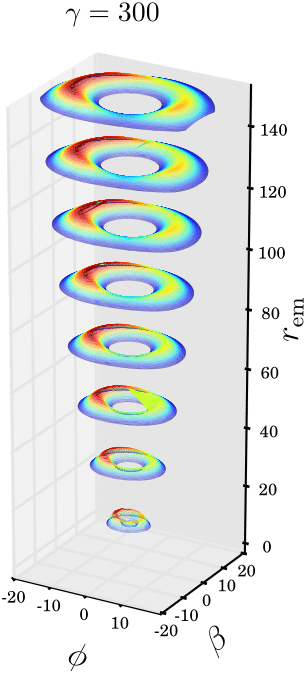}
\includegraphics[scale=0.35]{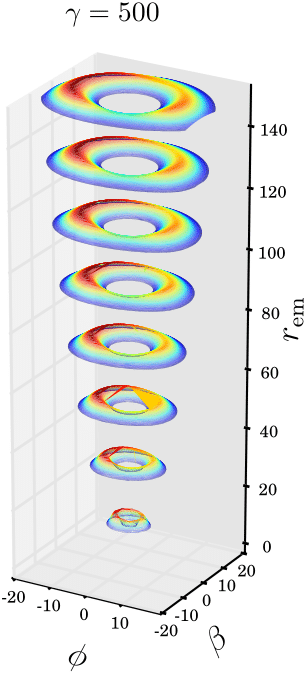}
\caption{Directivity pattern for various radiation altitudes for different average Lorentz factors $\gamma_0$. As it is discussed in Sect.~\ref{sec:dirpatt}, the absorption is amplified for lower $\gamma_0$ and the trailing peak is being damped.}
\label{fig:gammaDir}
\end{figure*}

In Figure~\ref{fig:j1022} we demonstrate this effect in simulated two-peaked pulsar (left panel) in comparison with real observational data for PSR~J1022+1001~\citep{daihobbs2015}. As we see, central hump can be easily reproduced as well. 


\begin{figure*}
\centering
\includegraphics[scale=0.35]{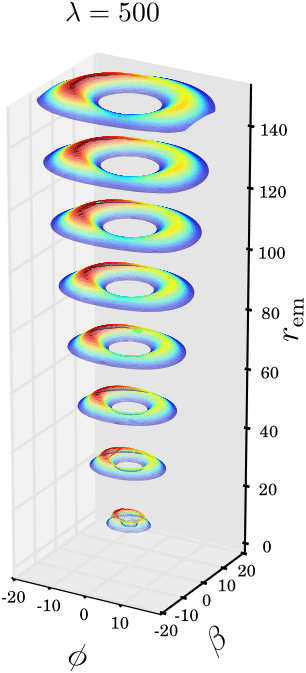}
\includegraphics[scale=0.35]{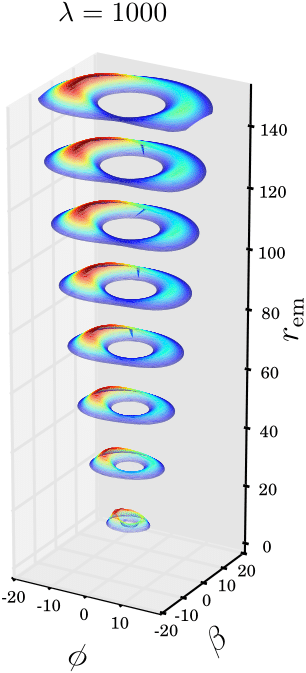}
\includegraphics[scale=0.35]{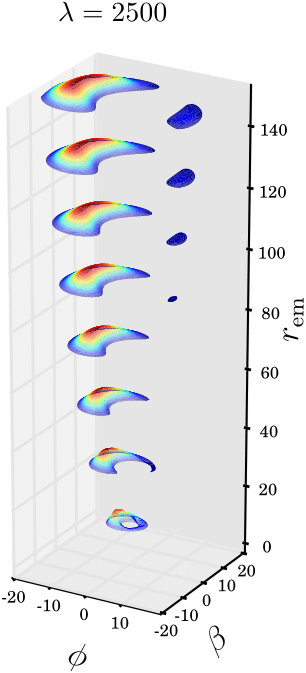}
\includegraphics[scale=0.35]{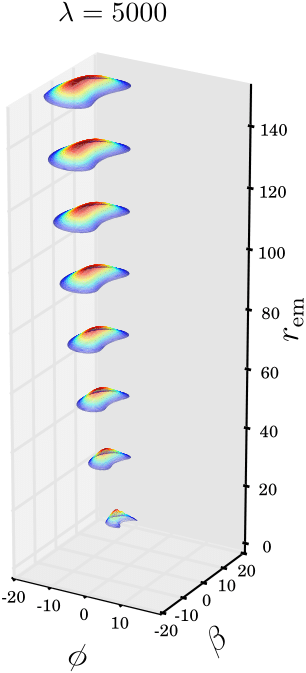}
\caption{Directivity pattern for various radiation altitudes for different plasma multiplicity $\lambda$. According to (\ref{eq:dsimple}) the higher the multiplicity the stronger the absorption, resulting in the complete disappearance of the trailing peak for high enough $\lambda$.}
\label{fig:lambdaDir}
\end{figure*}

\section{General properties}
\label{sec:genprop}

\subsection{Directivity pattern}
\label{sec:dirpatt}

At first, let us consider the effect of cyclotron absorption on the directivity pattern, i.e., the mean intensity of the profile for various emission radii $r_{\rm em}$. As was already shown, cyclotron absorption takes place in the region of weak magnetic field far away from the stellar surface, where the relation $\tilde{\omega}=\omega_B/\gamma$ holds~\citep{blandford76, mikhailovskii82, wanglaihan2010}. As in Paper I, we model the optical depth $\tau$ (\ref{eq:depth}) by the particle distribution function 
\begin{equation}\label{eq:distrib}
    F\left(\gamma\right)=\frac{6\gamma_0}{2^{1/6}\pi}\frac{\gamma^4}{(2\gamma^6+\gamma_0^6)},
\end{equation}
where $\gamma_0$ corresponds to mean Lorentz factor of secondary plasma. Such a distribution reproduces good enough the results of numerical simulations~\citep{dauharding82, BGI93}, e.g., power-law dependence $F(\gamma) \propto \gamma^{-2}$ for $\gamma \gg \gamma_{0}$.\par

As a result, two main plasma parameters affecting the strength of the resonance are the multiplicity $\lambda$ (\ref{eq:lambda}) and the mean Lorentz-factor $\gamma_0$. In Figure~\ref{fig:gammaDir} and Figure~\ref{fig:lambdaDir} we show the directivity pattern for various emission altitudes $r_{\rm em}$ (in star radii), where again $\phi$ is the phase and $\beta$ is the impact angle (minimum angle between the magnetic axis and the line of sight)\footnote{These patterns are well consistent with ones obtained by \citet{wangwanghan2014}, where propagation effects were taken into account as well.}. \par 

As was found in Paper I, the high multiplicity implies a strong absorption of the trailing peak. But the trailing peak reappears when we have high enough Lorentz factors. This is due to the fact, that the absorption radius $r_{\rm abs}\propto \gamma_{0}^{-1/3}$ (\ref{eq:rabs}) and hence $\tau\propto \lambda \gamma_0^{-1/3}$ (\ref{eq:dsimple}). In addition, this picture clearly shows, that the number of peaks of pulsar's profile is not a purely geometric property, but can also be a consequence of a strong synchrotron absorption. The only justified approach to distinguish between those two cases is to analyze the polarization curves.\par 


\begin{figure}
\centering
\includegraphics[scale=0.4]{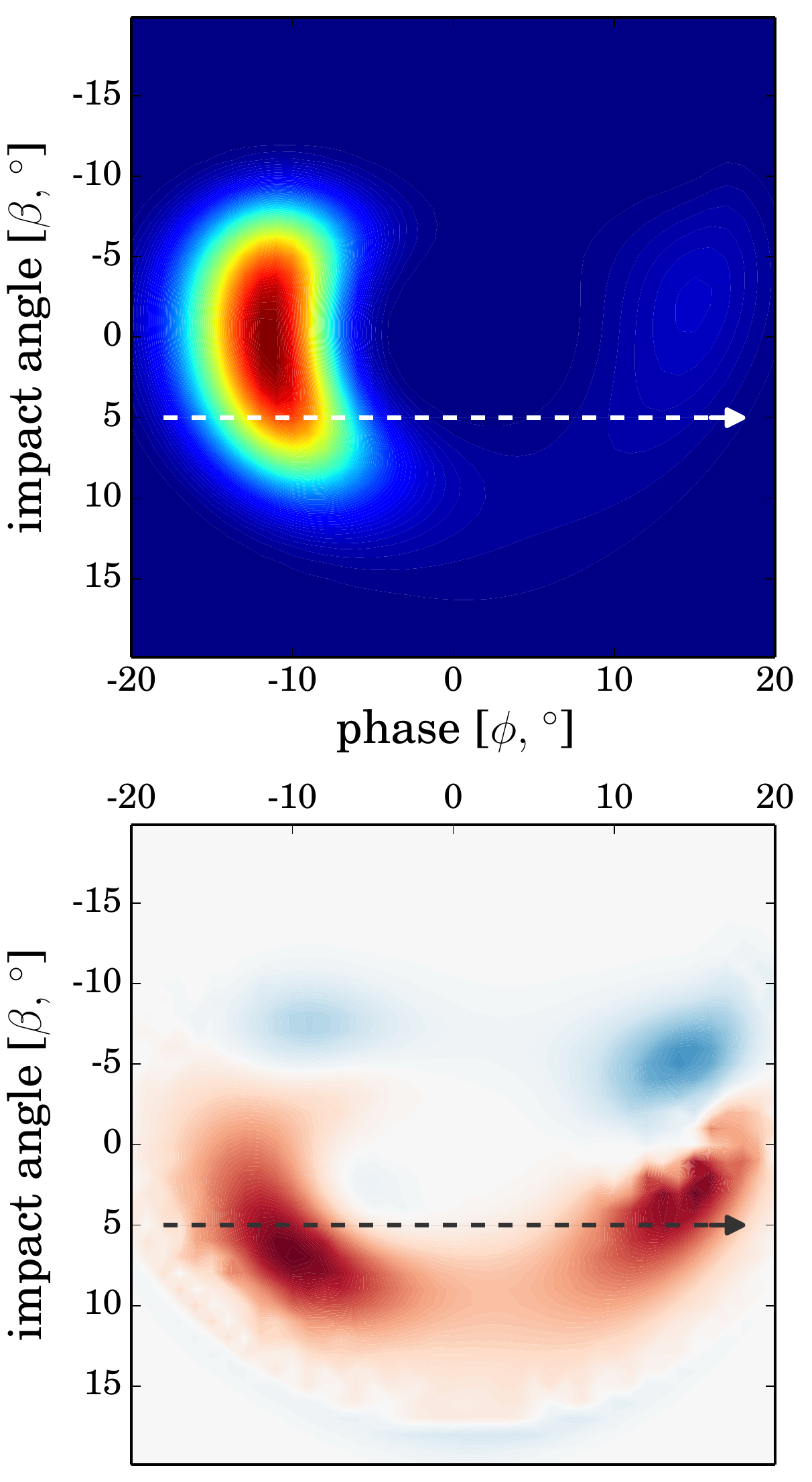}
\includegraphics[scale=0.5]{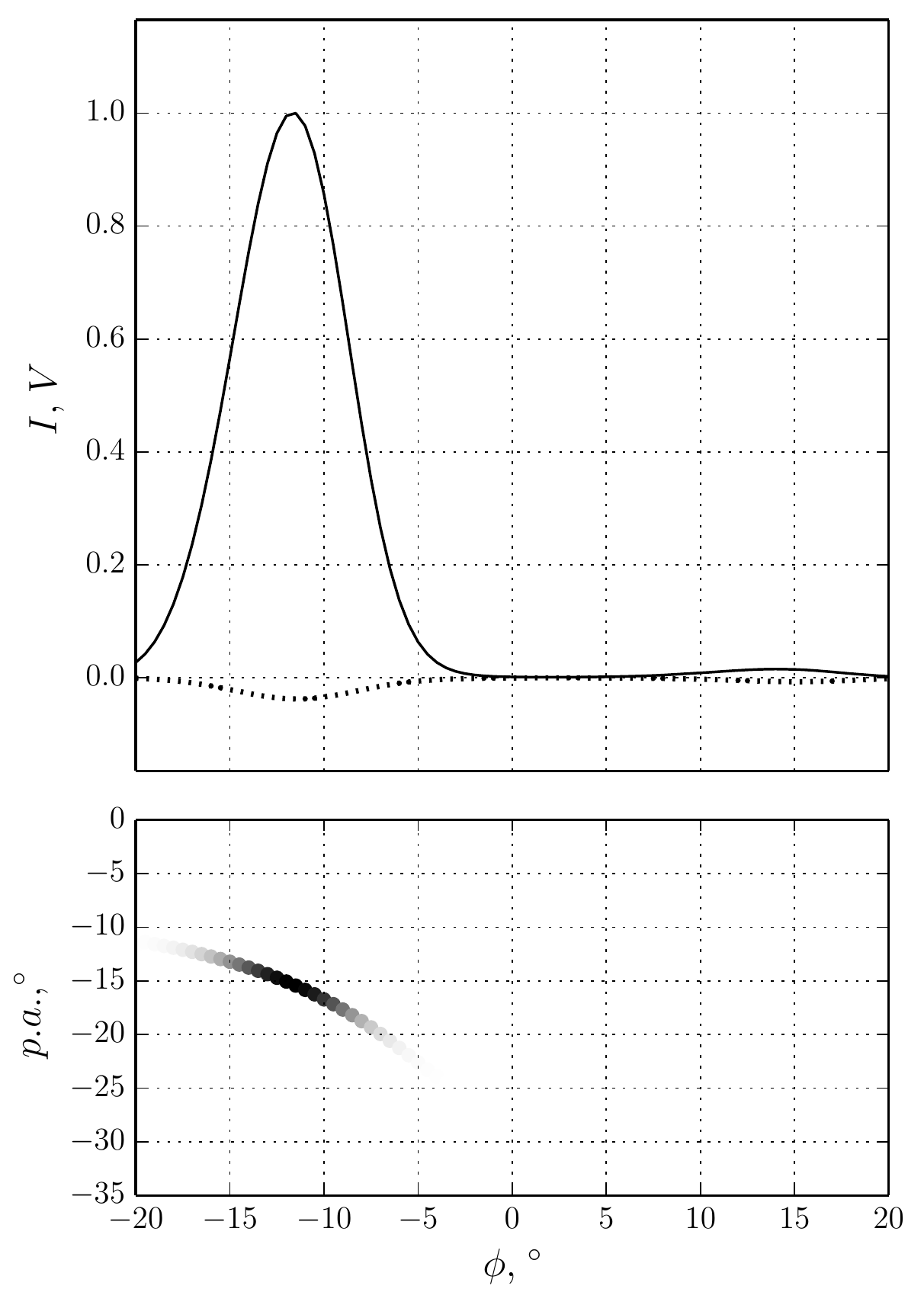}
\caption{Directivity pattern and circular polarization pattern at a radiation altitude. Single peaked pulsars for small impact angle ($\beta=5\degr$, $\lambda=5000$ and $\gamma_0=100$), dashed line represents the path of the line of sight. If the trailing peak is damped, one can form a single peak profile with a small impact angle, i.e., crossing the directivity pattern close to the center. The position angle is monotonous in this case, since the orientation of the magnetic field does not change over a single peak.}
\label{fig:sinpeak1}
\end{figure}


\begin{figure}
\centering
\includegraphics[scale=0.4]{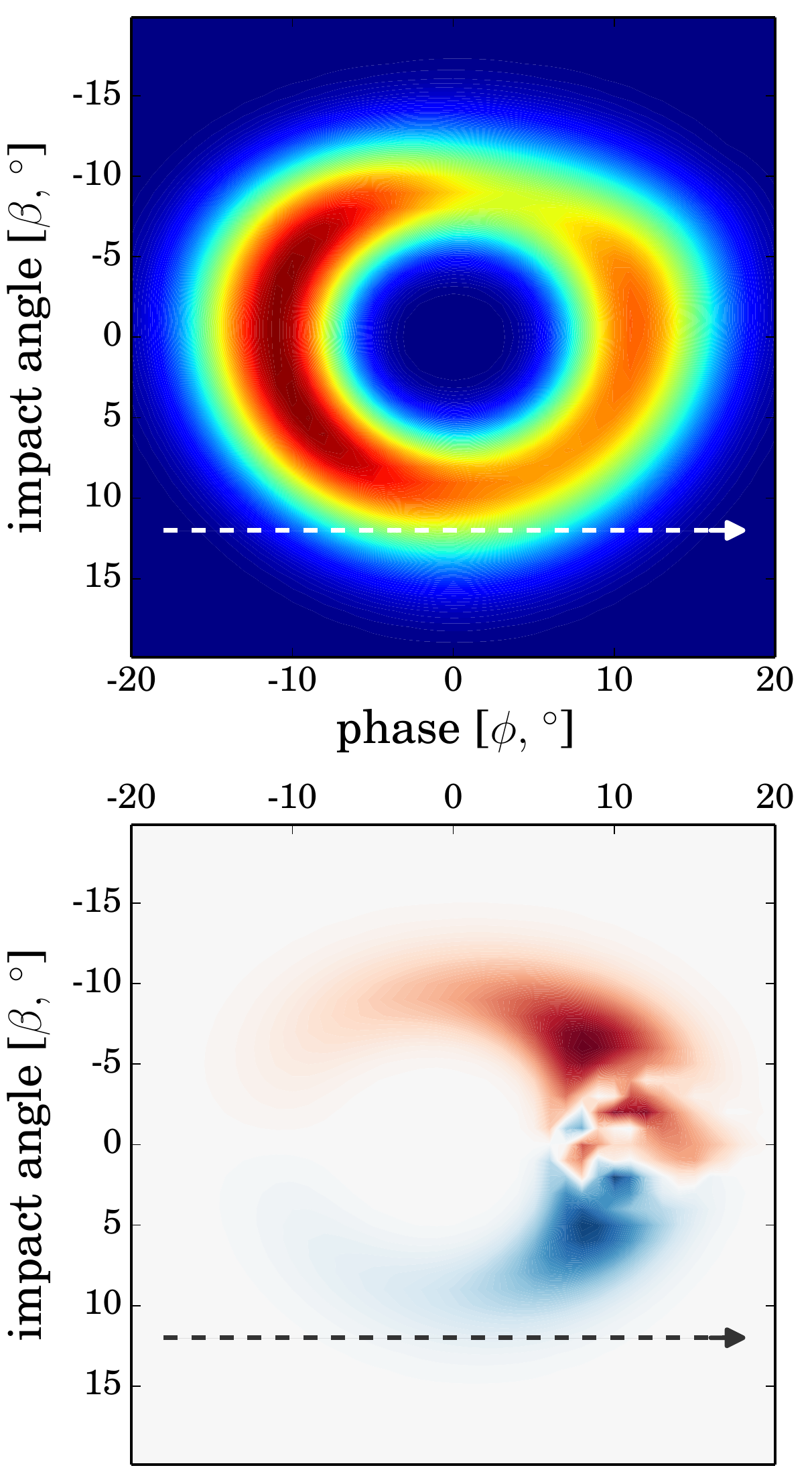}
\includegraphics[scale=0.5]{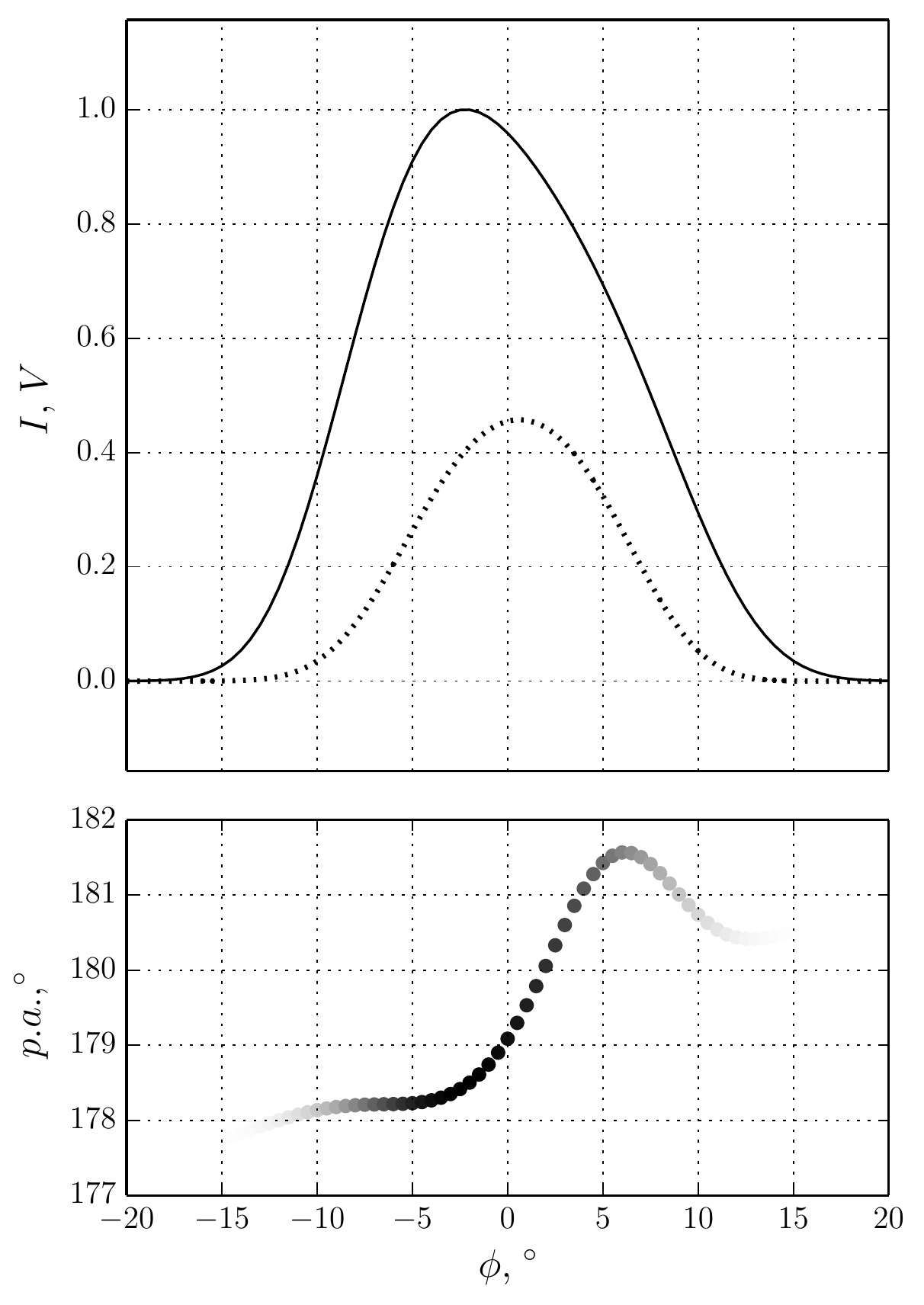}
\caption{Directivity pattern and circular polarization pattern at a radiation altitude. Single peaked pulsars for large impact angle ($\beta=12\degr$, $\lambda=1000$ and $\gamma_0=500$), dashed line represents the path of the line of sight. Another case of single peaked profile can be obtained by crossing the pattern near its edge. In this case the position angle undergoes a jump, due to a swipe in the magnetic field orientation along the peak.}
\label{fig:sinpeak2}
\end{figure}

In Figure~\ref{fig:sinpeak1} and Figure~\ref{fig:sinpeak2} we present two geometrically different cases, where one obtains single peak. In the first case, the line of sight crosses the directivity pattern near its center, but the trailing peak is suppressed due to high multiplicity $\lambda=5000$, resulting in the only peak. In the second case we have small multiplicity and large Lorentz factor, so the directivity pattern is a hollow circle. But now the line of sight crosses pattern near its boundary ($\beta=12\degr$).\par 

It is clear, that for the first case the $p.a.$ curve is to be smooth and even close to flat, as we cross just one part of the directivity pattern and the projection of the magnetic field onto the picture plane does not change strongly.  The pulsars J1739-3023 and J1709-4429 ~\citep{WJ2008} as well as B0656+14, B0950+08, and B1929+10~\citep{hankinsrankin2010} definitely belong to this class. However, if the position angle jumps as the line of sight passes through the center of the profile, as in J1224-6407, J1637-4553, J1731-4744, and J1824-1945~\citep{WJ2008}, one can be sure that we meet the second case. \par 

One can also note in Figure~\ref{fig:sinpeak1}, that the circular polarization level $V/I$ is much larger for the leading peak, than for the trailing one, although the resonant suppression is strong for the first one. This is due to the fact, that the escape height for the leading peak lies in the region of lower plasma density, than for the trailing one. 

\begin{figure}
\centering
\includegraphics[width=\columnwidth]{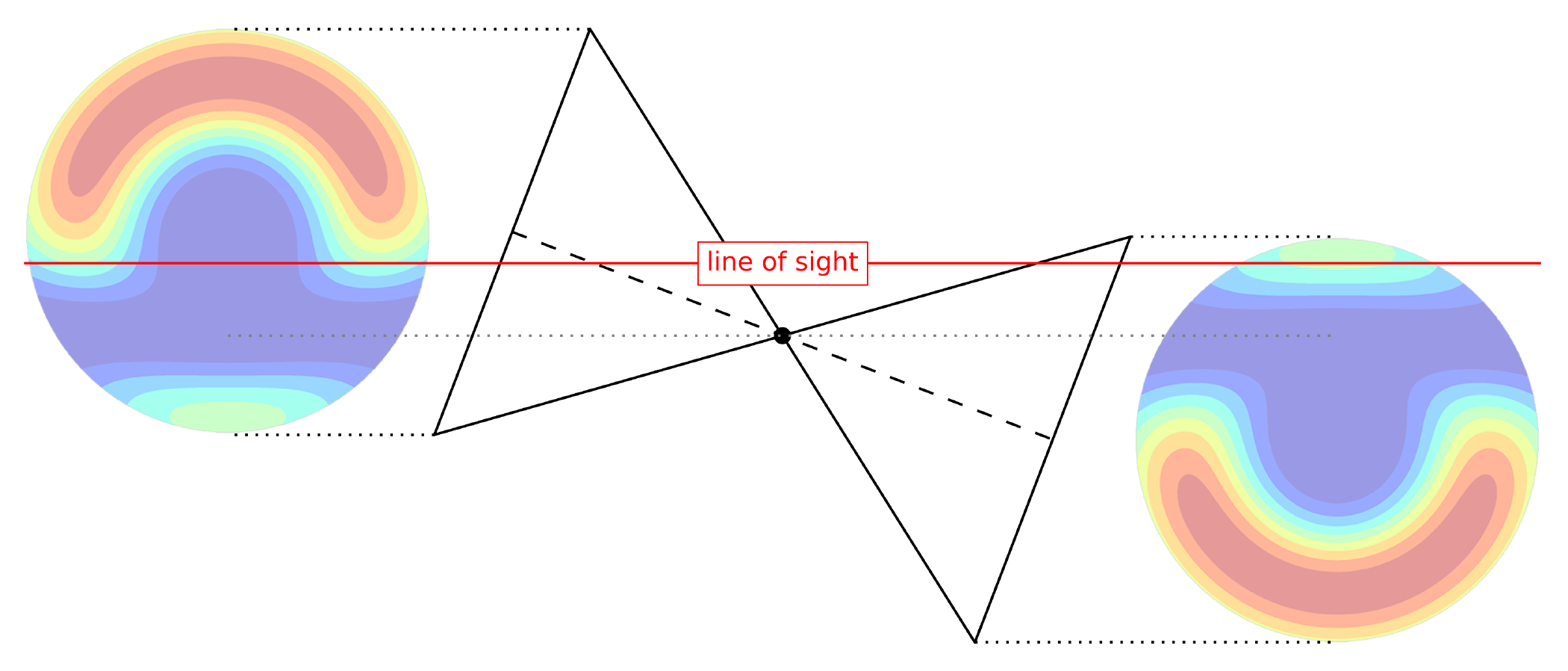}
\caption{The schematic illustration of pulsar with interpulse, radiating from opposite poles. The oblique neutron star (in the center) has two radiating cones (black lines). The line of sight (red line) crosses the hot radiating regions in the opposing poles, resulting in the main pulse and interpulse.}
\label{fig:interpulse}
\end{figure}

\subsection{Pulsars with interpulses}
\label{sec:interpulse}
 
In most cases the interpulses, i.e., distinct radiation features separated from the main pulse by the phase $\phi$ close to $180\degr$~\citep{manlyne, cadyritchings, kramerxilouris}, are thought to originate from the pulsar's opposite pole~\citep{maciesiakgil2011}. Thus, those pulsars are believed to have an inclination angle close to $90\degr$ and hence their analysis is important in the context of obliquity angle evolution \citep{WJ2008b} and the directivity pattern and polarization formation~\citep{keith2010}.\par


\begin{figure}
\centering
\includegraphics[scale=0.3]{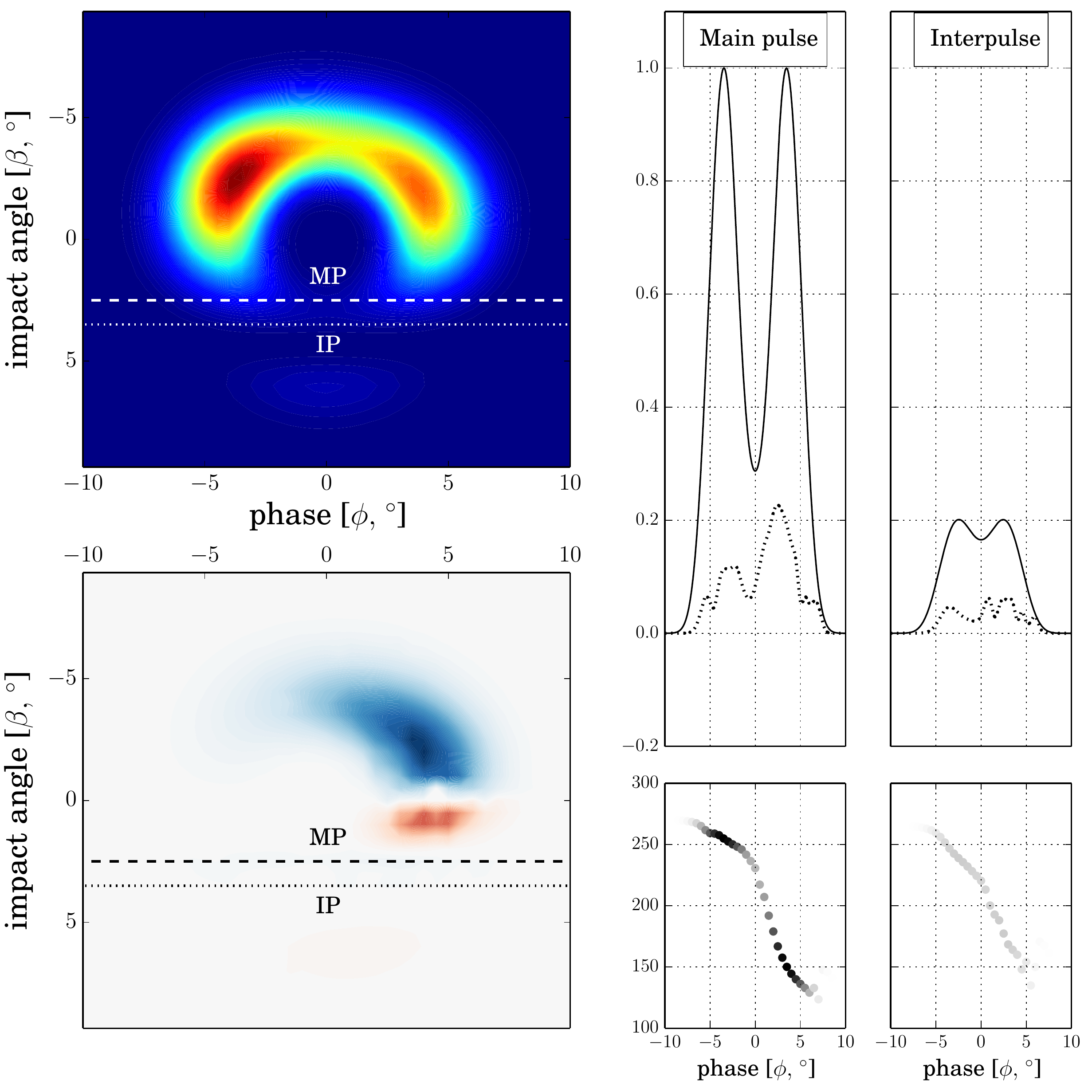}
\includegraphics[scale=0.3]{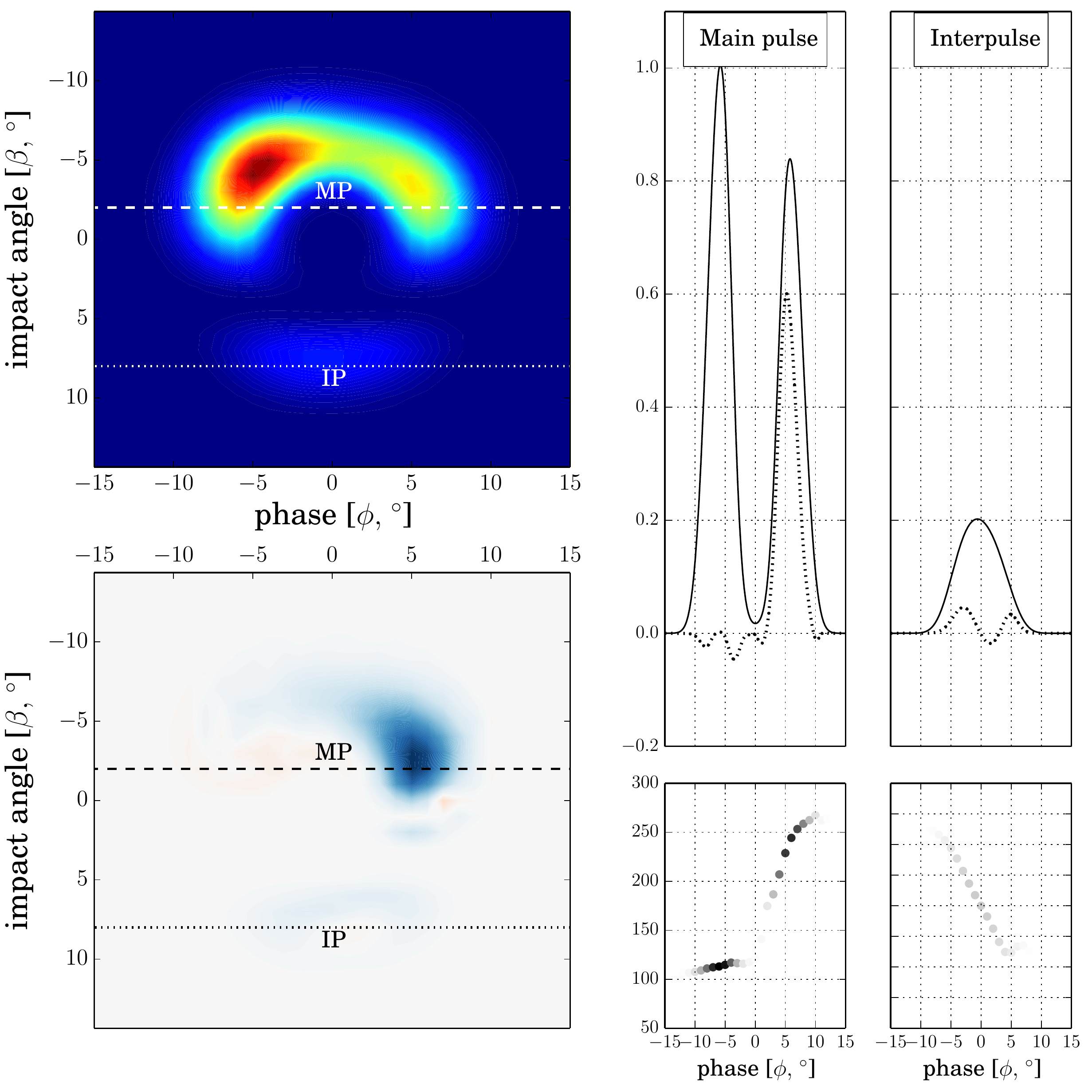}
\caption{Directivity pattern and circular polarization pattern at a radiation altitude. Inclination angle $\alpha\approx 85\degr$. If both the main pulse and interpulse are formed in the same (different) "hot" region of the pattern, the position angle curves will have similar (opposite) slopes.}
\label{fig:inter1}
\end{figure}

\begin{figure}
\centering
\includegraphics[scale=0.3]{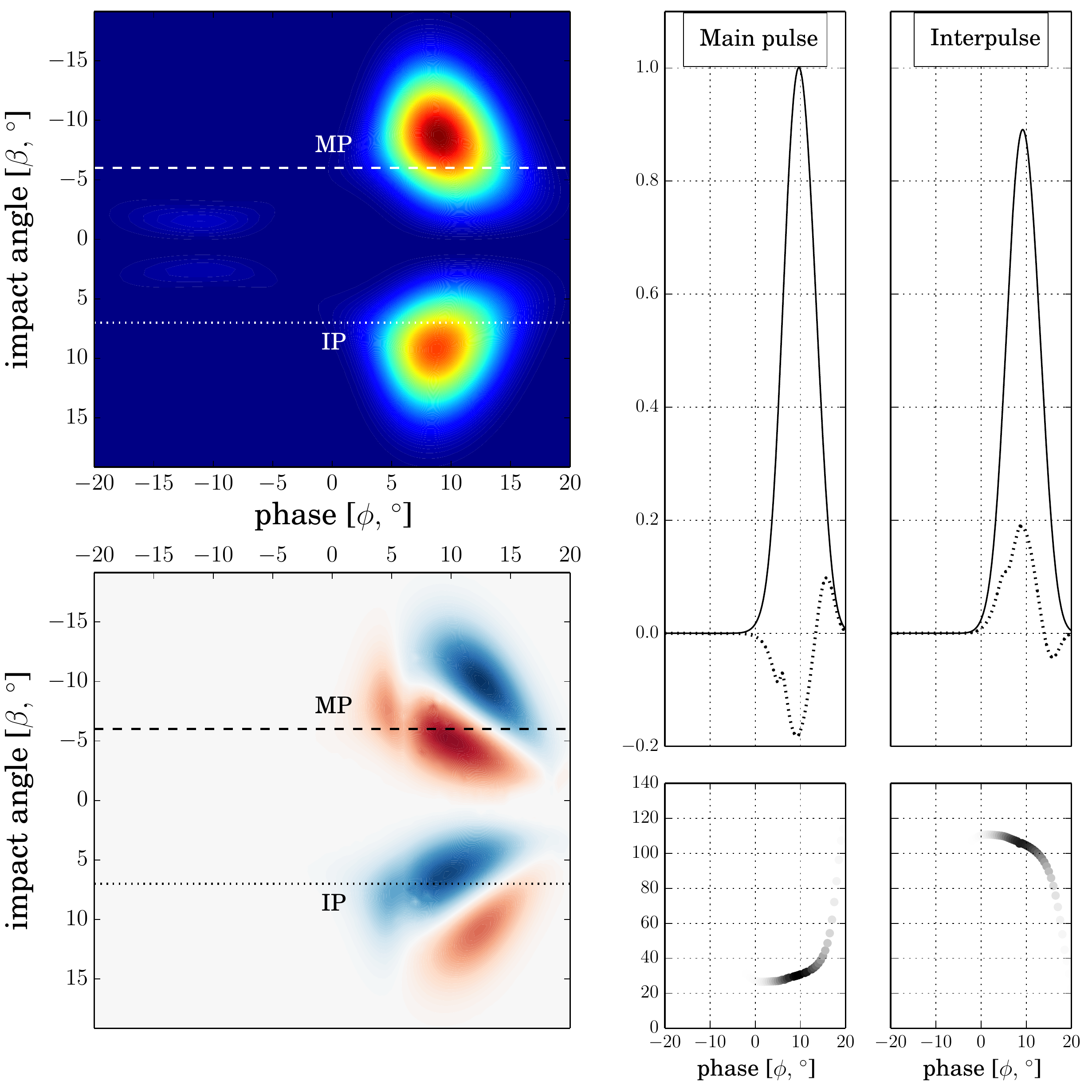}
\includegraphics[scale=0.3]{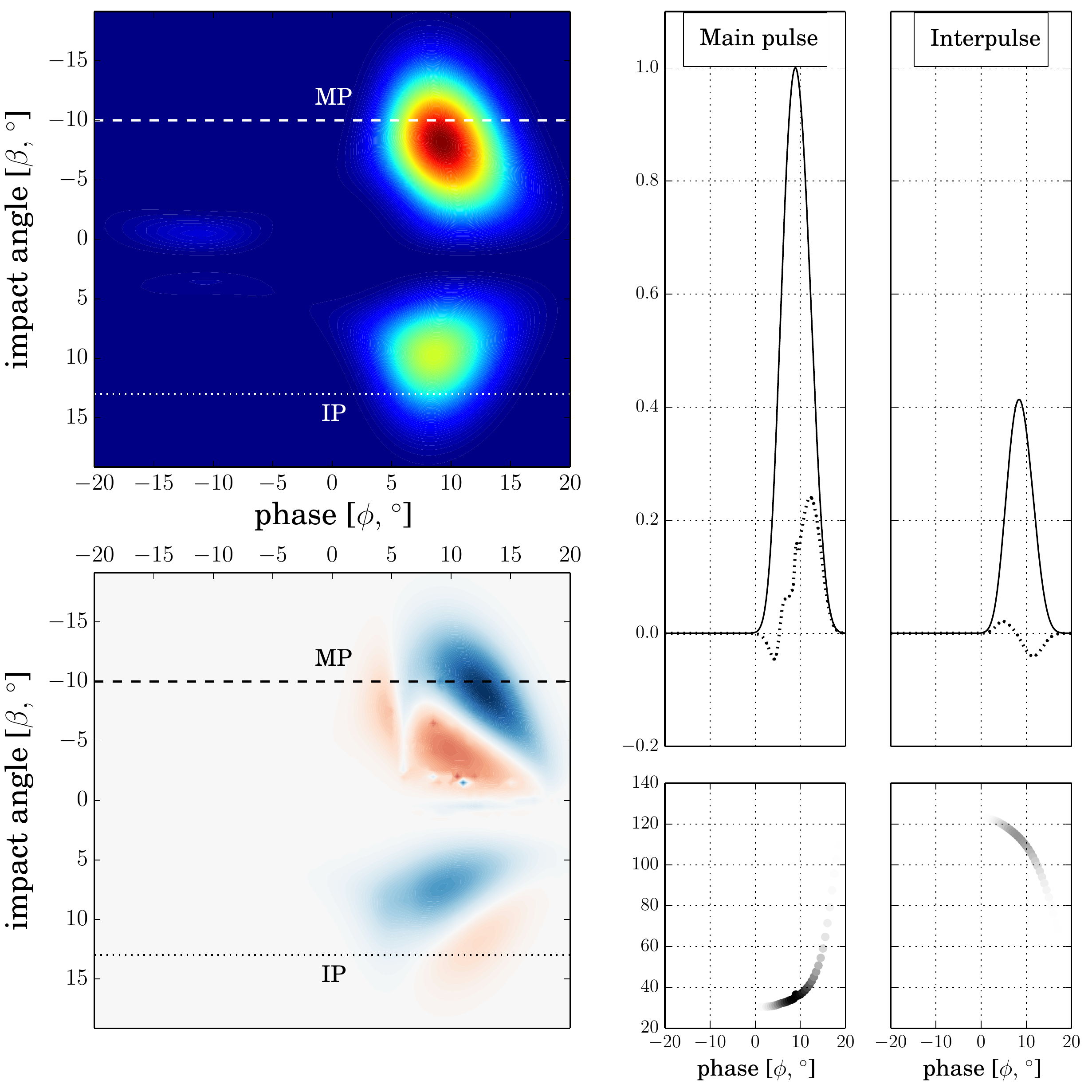}
\caption{Directivity pattern and circular polarization pattern at the radiation altitude. For inclination angles $\alpha\approx 90\degr$ the main pulse and interpulse are always formed in opposite regions of the pattern, resulting in opposite slopes of the p.a. and opposite circular polarizations. The MP and IP amplitudes ratio can be adjusted by varying the impact angle.}
\label{fig:inter2}
\end{figure}

Here we present the modeled profiles and polarization curves of pulsars with interpulses, both for the main pulse (MP) and the interpulse (IP). As the line of sight crosses actually the same directivity pattern with two different impact angles (see Figure~\ref{fig:interpulse}), we are able to model both the MP and the IP on the same directivity pattern. In Figure~\ref{fig:inter1}-\ref{fig:inter2} the two cases are presented with different geometric parameters. The perturbations in the directivity pattern, given by the density profile (\ref{eq:profile}), cause the formation of two emitting regions, separated by the suppressed density gap on the magnetic field lines intersecting neutron star surface where the condition $\bmath{\Omega}\cdot\bmath{B}\approx 0$ is satisfied. Dashed and dotted lines represent respectively the line of sight path for the main pulse and the interpulse.\par

In Figure~\ref{fig:inter1} we demonstrate the case $\alpha\approx 85\degr$, while in Figure~\ref{fig:inter2} the orthogonal geometry $\alpha\approx 90\degr$ is presented. As one can see, there are clearly two distinct "hot" regions, and while in the second case they are symmetric, generating IP with roughly the same amplitude ($>50\%$), in the first case the lower region is suppressed, as it is located in the rarified plasma region, and this results in a large difference in the intensity of the MP and IP ($10-30\%$). In the first two pictures (Figure~\ref{fig:inter1}) we show the MP and IP generating along the same region (corresponding impact angle $\beta=3\degr$), which results in roughly the same $p.a.$ curve and similar circular polarization level (see, e.g., PSR~J1722-3712). On the other hand, for $\beta=-2\degr$ the MP and IP are generated in different regions of the directivity pattern, having the opposite run of the position angle (see, e.g., PSR~J1549-4848). In the second case (the leading part of the directivity pattern is suppressed due to large $\lambda$ (see Sect.~\ref{sec:dirpatt}), as the inclination angle is close to $90\degr$, MP and IP always cross the opposite "hot" regions of the directivity pattern (see Figure~\ref{fig:inter2}). As these regions are close to symmetric, we end up having similar amplitudes for MP and IP and opposite run of the $p.a.$ and circular polarization.\par

\subsection{Width of the emitting region}
\label{sec:radtofreq}

Understanding the size and width of the region where the radio emission originates is a crucial step towards a construction of a self-consistent radiation theory. Although there are no direct methods of determining the actual altitude of that region, there are several naive approaches that may help to do rough estimates. Namely, geometrical 'hollow cone' model together with the A/R effects~\citep{blaskiewicz91, mitragupta2009} showed that the radiation can originate in the region from $10$ to $100$ stellar radius. However, this approach is not physically-motivated if propagation effects are significant.\par


\begin{figure*}
\centering
\includegraphics[scale=0.55]{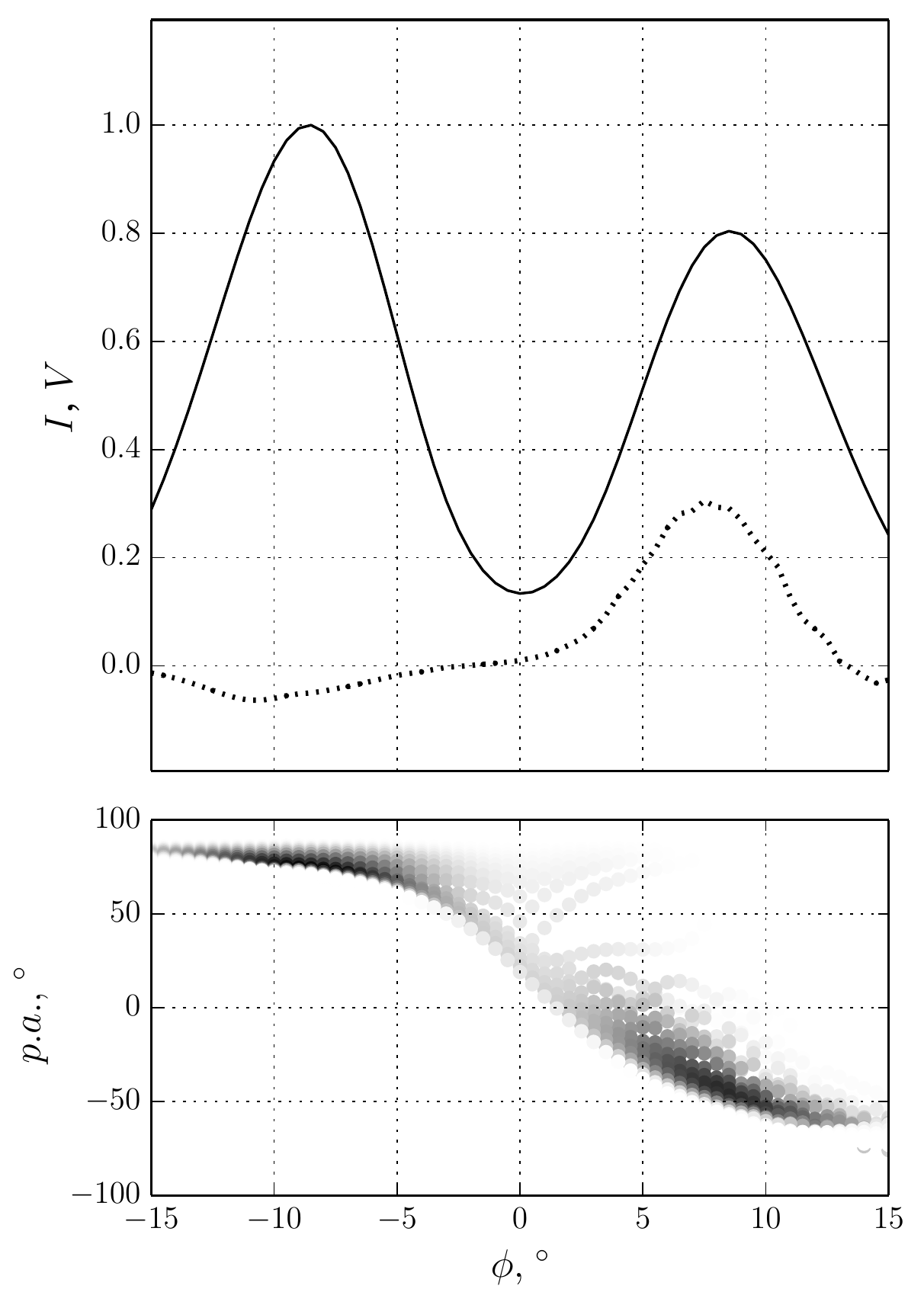}
\includegraphics[scale=0.55]{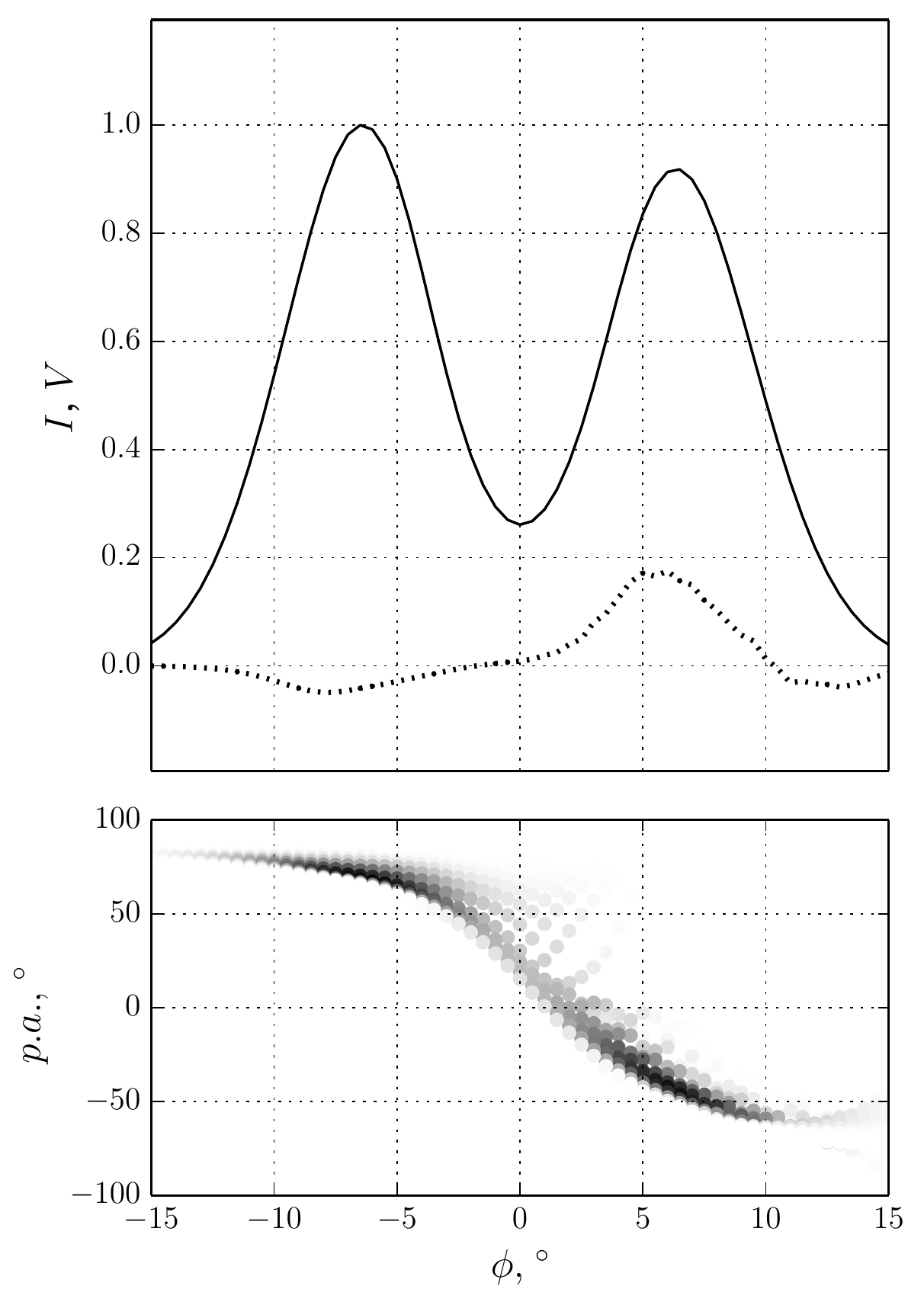}
\caption{Simulated profiles for O-mode pulsar PSR~B0301+19 at two distinct frequencies 430 MHz (left) and 1414 MHz (right). The scattering is due to the generation on a wide range of altitudes. For both cases the position angle curve width is larger near the edges, since the intensity in the center is suppressed. However, since higher frequencies are thought to be generated closer to the stellar surface, the characteristic p.a. curve width for the second case is smaller.}
\label{fig:b0301}
\end{figure*}

\begin{figure*}
\centering
\includegraphics[scale=0.38]{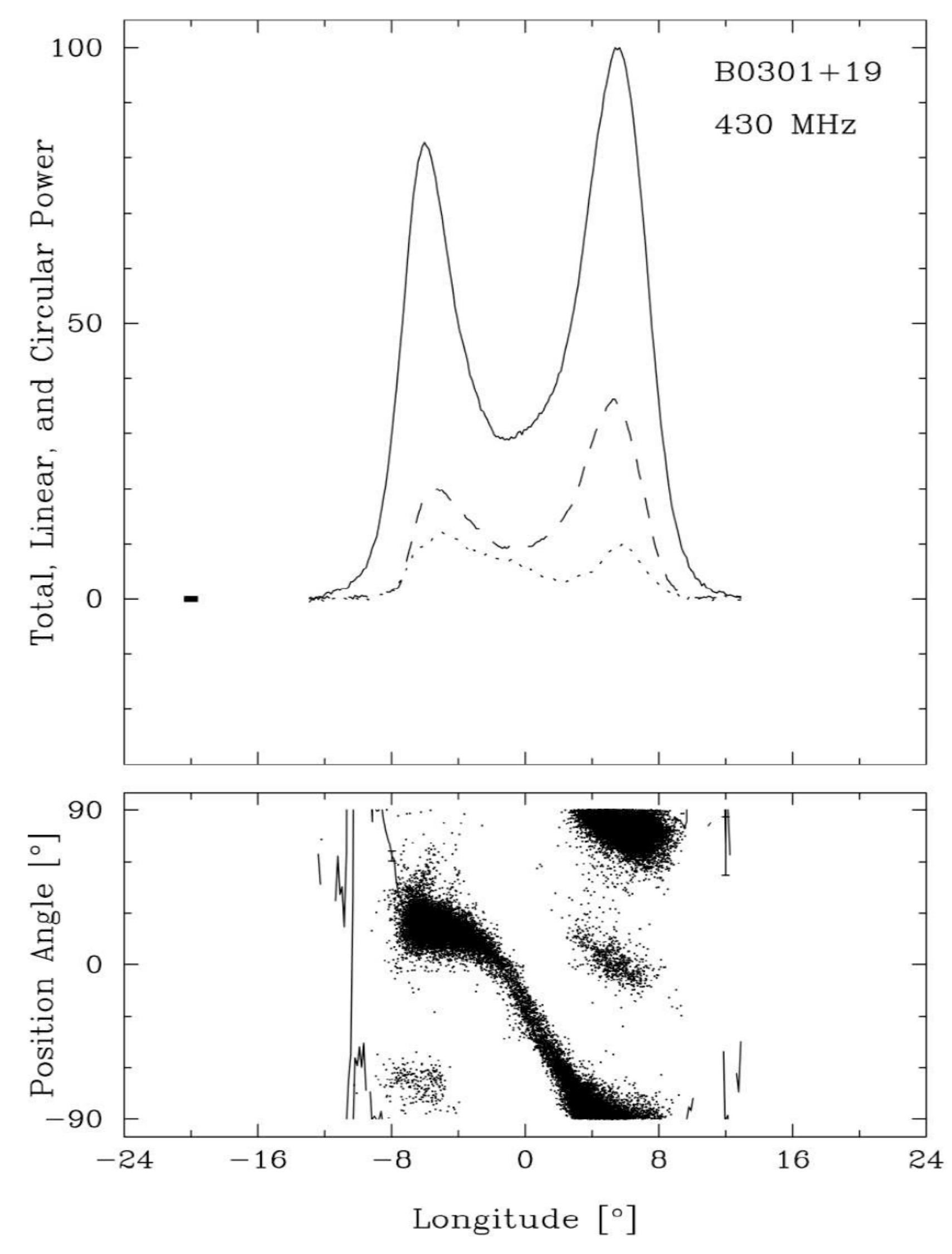}
\hspace{1em}
\includegraphics[scale=0.38]{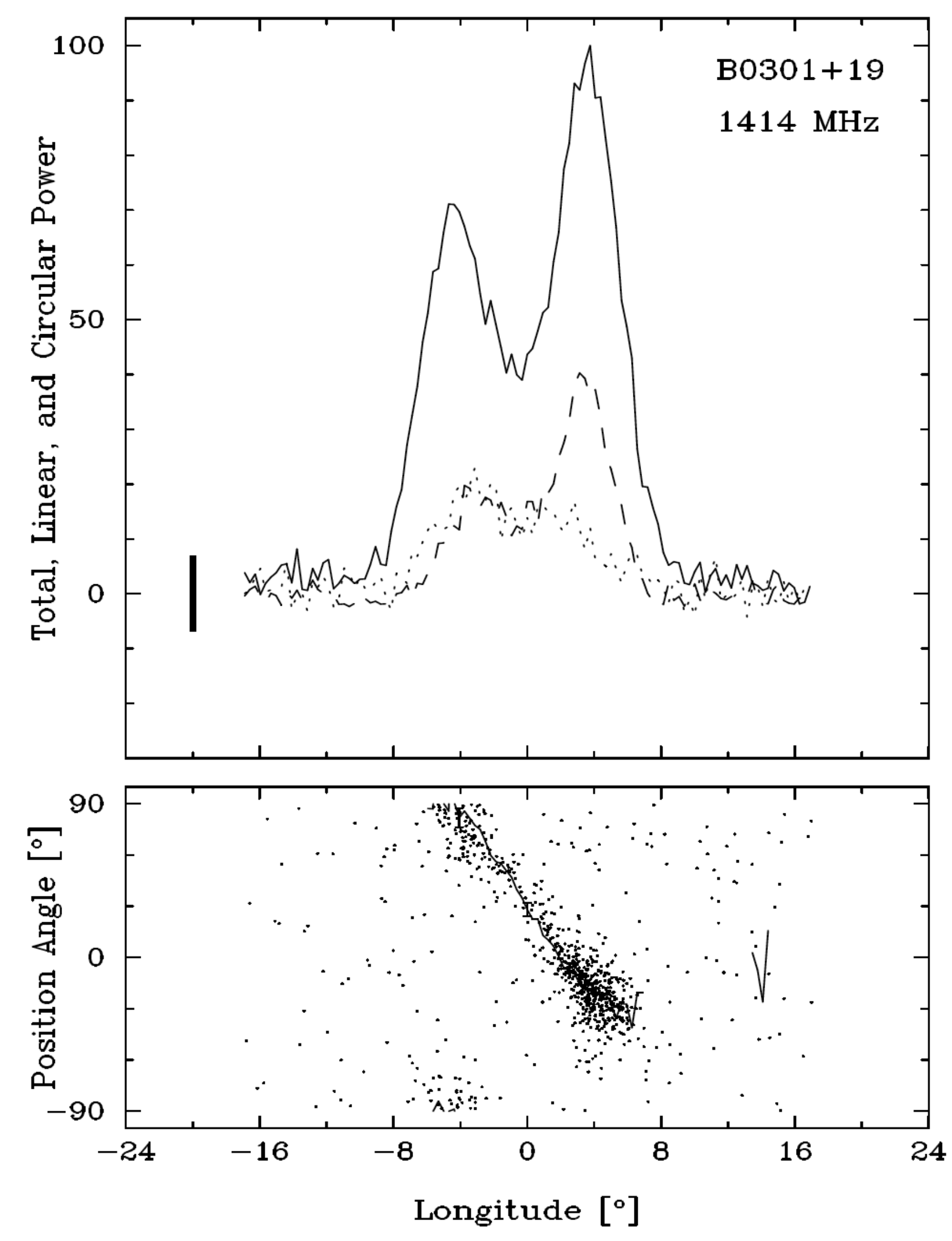}
\caption{Observed profiles for O-mode pulsar PSR~B0301+19 at two distinct frequencies (images taken from \citealt{hankinsrankin2010}).}
\label{fig:b0301o}
\end{figure*}

In this paper we propose a method of conducting the radius-to-frequency mapping that allows us to evaluate the height and characteristic depth of the radiation region using polarization characteristics. To do that, we compare the results of our simulation of the $p.a.$ with the corresponding observational plots obtained by~\citet{hankinsrankin2010} who presented the $p.a.$ curve with characteristic distinct scatter points. Such a scattering  can be explained if we assume that the radiation originates not from one particular radius, but from a rather wide shell.\par 


\begin{figure*}
\centering
\includegraphics[scale=0.55]{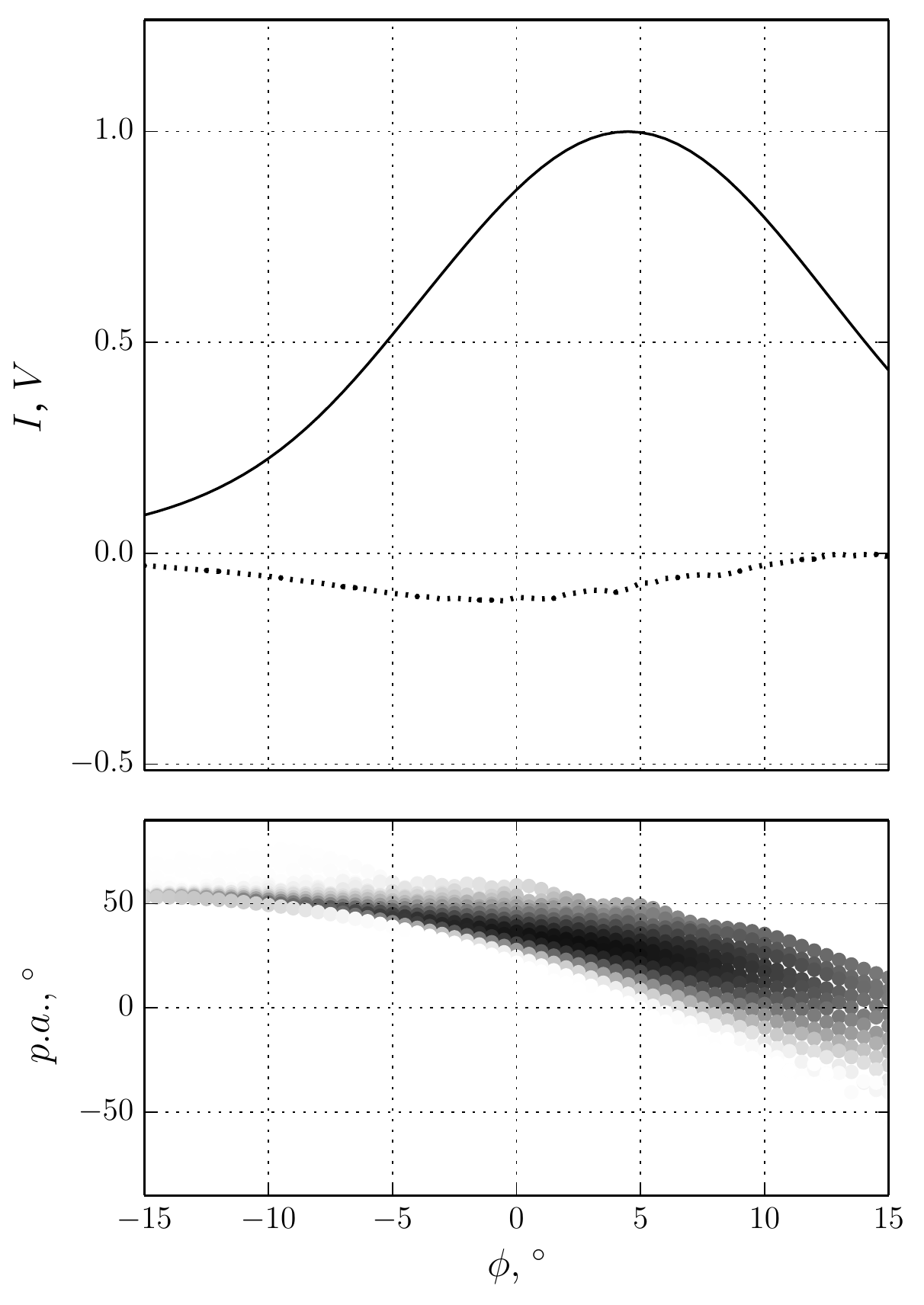}
\includegraphics[scale=0.55]{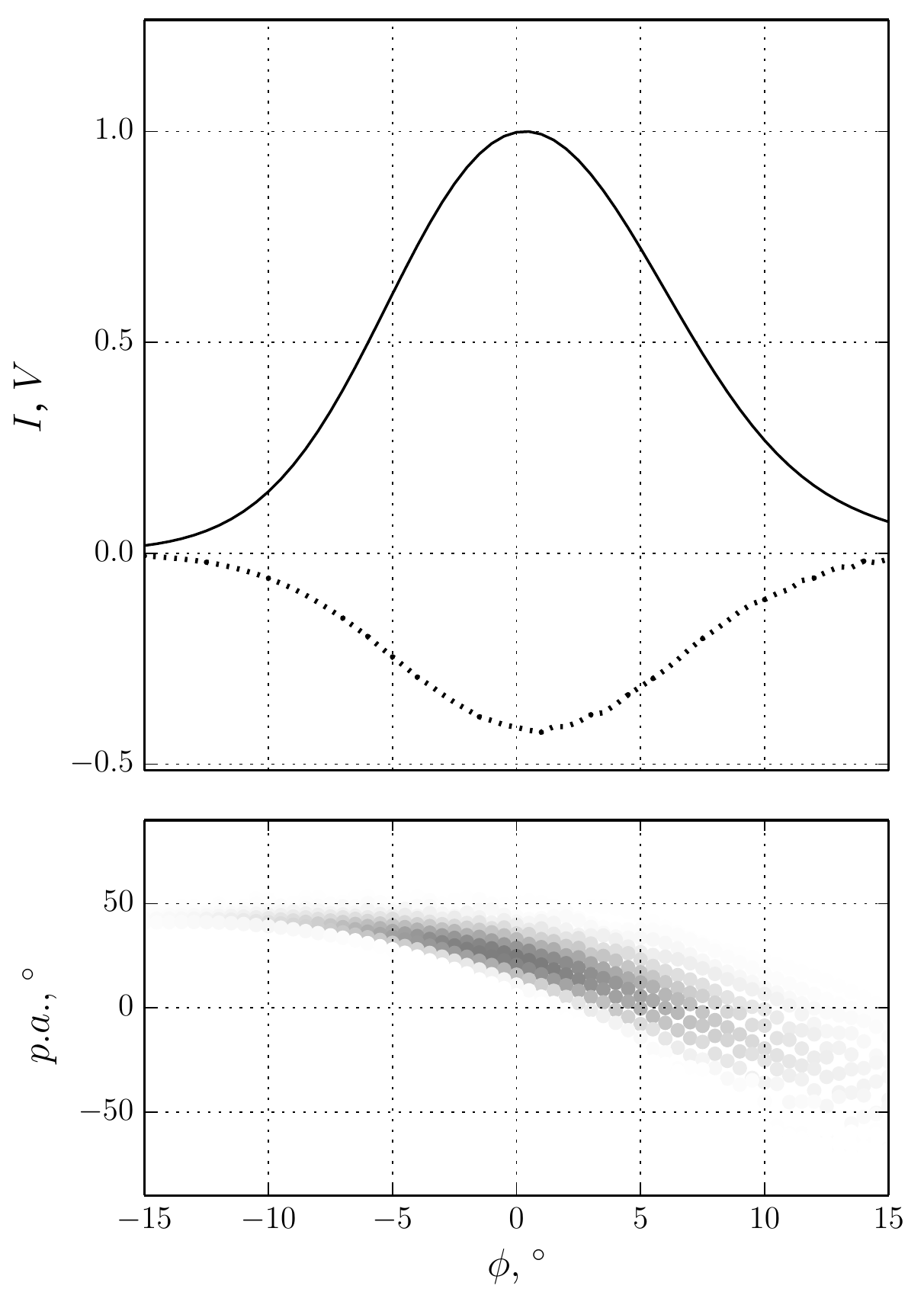}
\caption{Simulated profiles for X-mode pulsar PSR~B0540+23 at two distinct frequencies 430 MHz (left) and 1414 MHz (right). The scattering is due to the generation on a wide range of altitudes. For both cases the position angle curve width is smaller near the edges, since the intensity is suppressed there. However, since higher frequencies are thought to be generated closer to the stellar surface, the characteristic p.a. curve width for the second case is smaller.}
\label{fig:b0540}
\end{figure*}

\begin{figure*}
\centering
\includegraphics[scale=0.38]{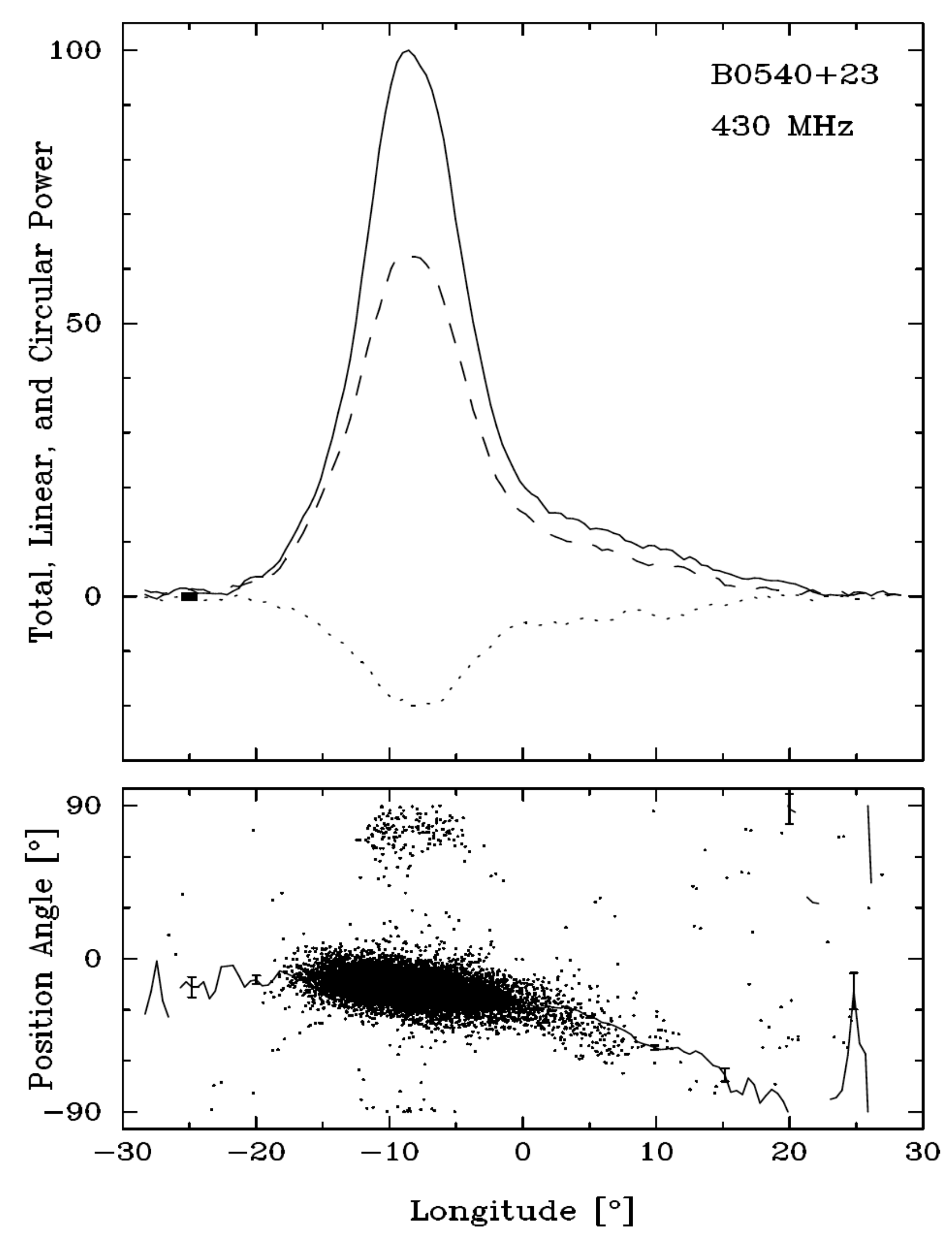}
\hspace{1em}
\includegraphics[scale=0.38]{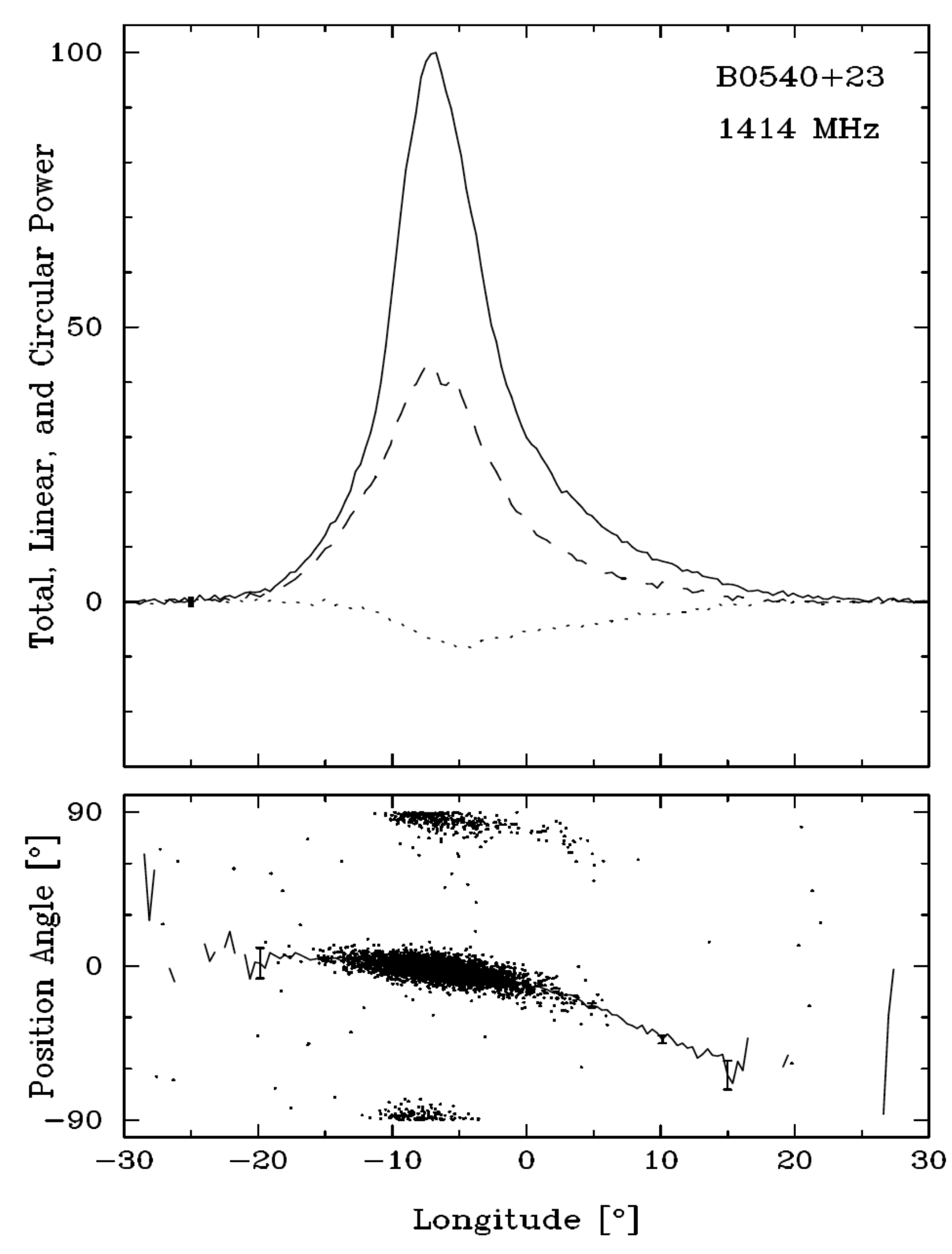}
\caption{Observed profiles for X-mode pulsar  PSR~B0540+23 at two distinct frequencies (images taken from \citealt{hankinsrankin2010}).}
\label{fig:b0540o}
\end{figure*}

In Figure~\ref{fig:b0301} we present the results of such analysis for the two-peaked pulsar PSR~B0301+19 compared to the observational curves presented in Figure~\ref{fig:b0301o}. We approximated the scatter curve with the parameters shown in Table~\ref{tab:b0301} where $\Delta p.a.$ is the rough scatter dispersion of the position angle data points. As we see, for double-peaked mean profile (which we connect with the O-mode) the width of the p.a. curve is slimmer in the center of a profile and wider near the pulse edges. This common property which is observed in all double-peak O-mode pulsars can be easily explained. Indeed, as is shown on Figures~\ref{fig:gammaDir} and~\ref{fig:lambdaDir}, the central 'hole' of the directivity pattern increases in size with the generation radius $r_{\rm em}$. Thus, only the very deep parts of the radiation domain give the observable radiation in the central part of the mean pulse. As to pulse edges, they will be radiated from all the generation domain. In addition, as for higher frequency the thickness of the $p.a.$ curve (which is directly corresponded to the radiation region size) is smaller, one can conclude that the radiating shell is smaller as well, which is due to the fact that higher frequencies are generated in the deep regions close to the stellar surface. \par

\begin{table}
\caption{Radiation regions for PSR~B0301+19 and PSR~B0540+23. Here $\Delta p.a.$ is the rough scatter dispersion of the position angle data points taken from~\citet{hankinsrankin2010}}
\label{tab:b0301}
 \begin{tabular}{ccccc}
 \hline
 PSR & $\nu$, MHz & $\Delta p.a.$ & $r_{\rm em}/R$ & $\Delta r_{\rm em}/R$ \\
  \hline
  \hline
 B0301+19 & 430 & $40\degr$ & 80 & 50 \\ 
 \hline
 B0301+19 & 1414 & $30\degr$ & 50 & 30 \\
 \hline
 B0540+23 & 430 & $50\degr$ & 70 & 80 \\ 
 \hline
 B0540+23 & 1414  & $30\degr$ & 40 & 30 \\
\hline
\end{tabular}
\end{table}

On the other hand, for the single-peaked X-mode pulsar PSR~B0540+23 (Figure~\ref{fig:b0540}) the $p.a.$ curve is wider in the center of integrated profile (due to the intensity suppression near the edges), which is also in a good agreement with observational data presented on Figure~\ref{fig:b0540o}. The estimated upper boundaries for the altitudes in this case are also presented in Table~\ref{tab:b0301}. Basically the same trend holds here: higher frequencies are generated on lower altitudes and have a narrower radiation region. This fact, however, does not appear to be universal as for some pulsars the higher frequencies may have a wider radiation region (that can be estimated from $\Delta p.a.$), while still originating from the deep altitudes (e.g., PSR~B0943+10, PSR~B1133+16, and PSR~B2020+28). \par

As a result, the key parameters, i.e., the inclination angle $\alpha$ and the impact angle $\beta$, as well as the approximate multiplicity parameter$\lambda$, the characteristic gamma-factor $\gamma_0$ and the radiation height $r_{\rm em}$ can be determined from the mean profile shape and circular polarization V. Those approximate values can further be corrected in the comparison with polarization data. After that we are left with only one parameter: $\Delta r_{\rm em}$, i.e., the characteristic depth of that region. In Table~\ref{tab:param} we present the parameters of the above mentioned pulsars that were used in our simulations. This estimations are rough and are based exclusively on mean profile $I$, circular polarization $V$ and position angle $p.a.$ curves that we compared with the observations. The period $P$ and magnetic field $B_{12} = B_{0}/(10^{12} \, {\rm G})$ was taken from \citet{hobbslyne2004} and \citet{wangmanchester2001}. \par

\begin{table}
\caption{Estimated pulsar parameters for $\lambda = 10^{3}$}
\label{tab:param}
\centering
\begin{tabular}{c c c c c c c}
 \hline
 PSR & $P$, s & $B_{12}$ & $\gamma_0$ &  $\alpha(\beta)$ & $R_{\rm L}/R$ & $r_{\rm esc}/R$\\
 \hline\hline
 B0301+19 & $1.39$ & $1.4$ &  $200$ &  $60\degr(3\degr)$ & $6700$ & $300$ \\ 
 \hline
 B0540+23 & $0.25$ & $2.0$ &  $100$ &  $60\degr(8\degr)$ & $1200$ & $1100$ \\ 
 \hline
 J2048-1616 & $1.96$ & $4.7$ & $400$ &  $60\degr(3\degr)$ & $9300$ & $200$ \\ 
 \hline
 J0738-4042 & $0.37$ & $0.8$ & $100$ & $50\degr(-16\degr)$ & $1800$ & $450$ \\
 \hline
\end{tabular}
\end{table}

To conclude, one can say that our approach, together with the observational scatter data for $p.a.$ (such as in catalog by \citealt{hankinsrankin2010}) provides a strong instrument for estimating the upper bounds for the radiation region altitudes.


\begin{figure*}
\centering
\includegraphics[width=2\columnwidth]{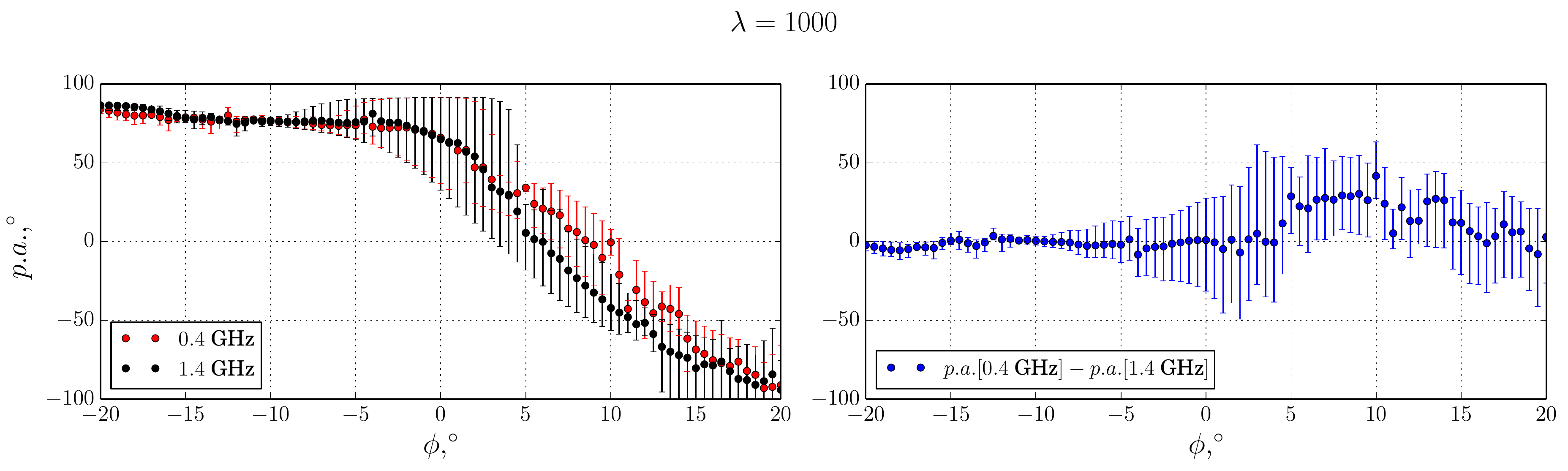}
\includegraphics[width=2\columnwidth]{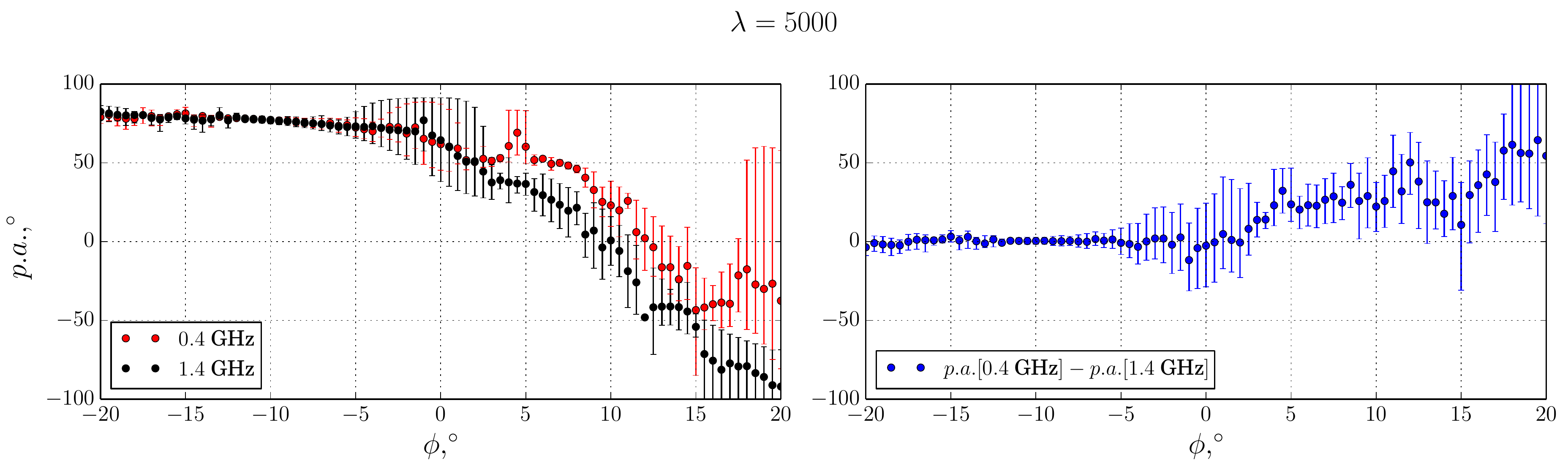}
\includegraphics[width=2\columnwidth]{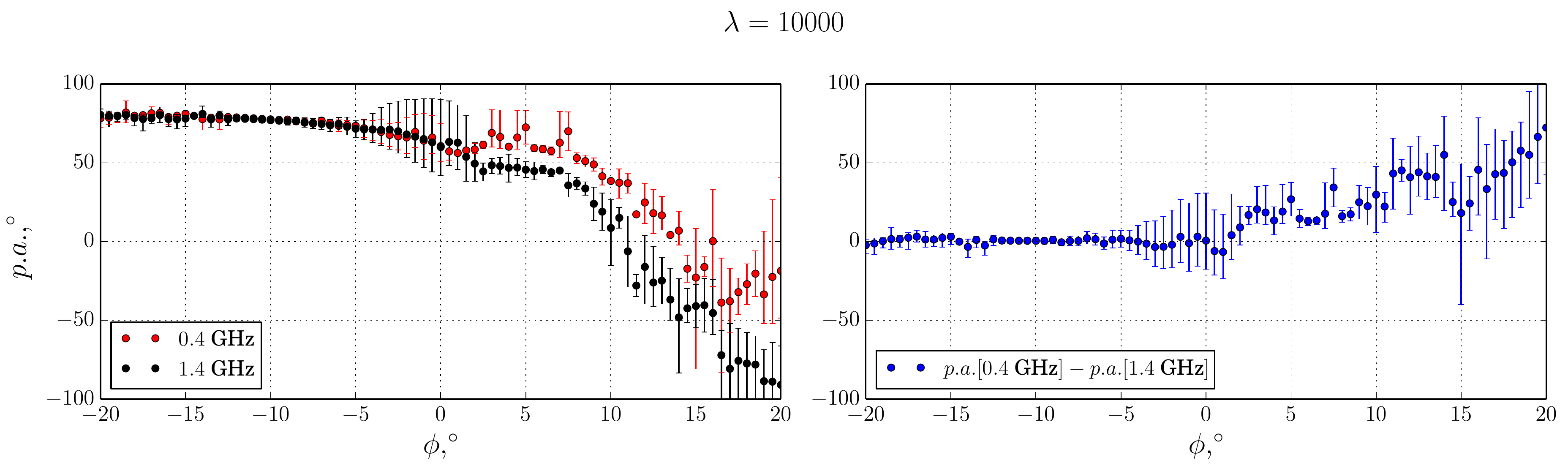}
\includegraphics[width=2\columnwidth]{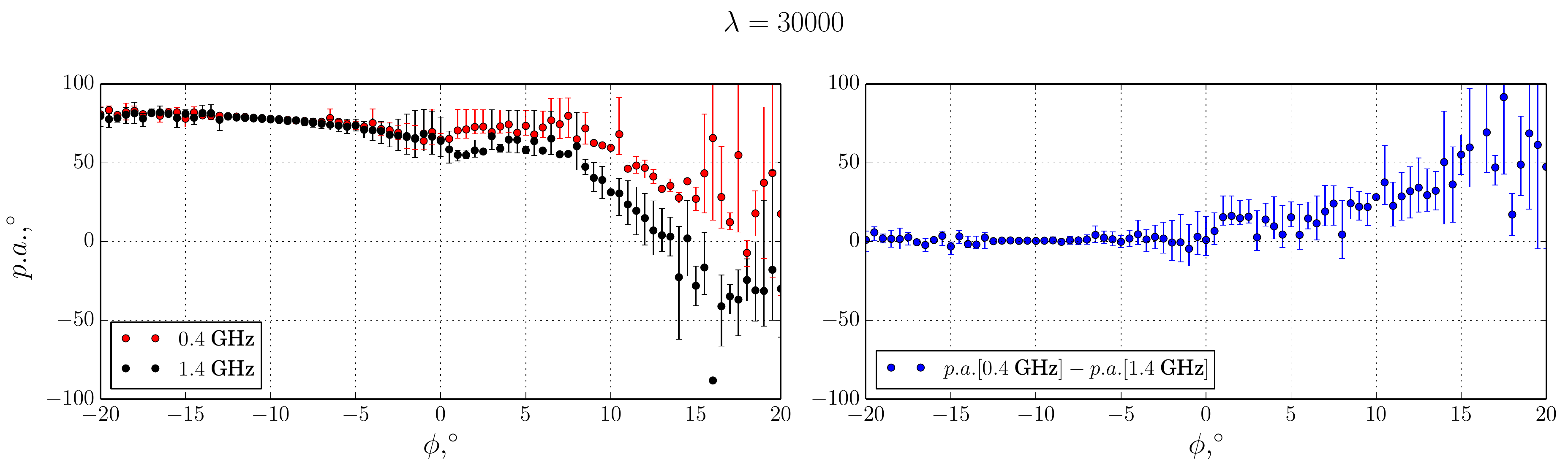}
\caption{Position angle curves on two various frequencies and their difference for different multiplicity. The error bars are modeled by taking a wide range of radiation altitudes. One can see, that in fact the p.a. curves can be close and overlap for two different frequencies and a wide range of multiplicities. This fact means, that propagation effects do not necessarily imply a strong dependence of the position angle curve on the frequency.}
\label{fig:pashift}
\end{figure*}

\subsection{Position angle shift}
\label{sec:posangshift}

The maximum of the $p.a.$ derivative $(\mathrm{d}p.a./\mathrm{d}\phi)_{\rm max}$, i.e., the center of the $p.a.$ curve, is shifted to the right relative to the center of the profile. This is a well known observational effect and it was a subject of study for a long time: see, e.g., PSR J0729-1448, J0742-2822, and J1105-6107 from \citet{WJ2008} and  J0631+1036, J0659+1414, J0729–1448,  J0742–2822, J0908–4913, and J1057–5226 from~\citet{Rookyard15}. As it was mentioned above, usually the $p.a.$ shift is assumed to be the consequence of the A/R effects. In this case, the position angle shift can be estimated as $\Delta \phi_{p.a.}\approx4r_{\rm em}\Omega/c$, and this dependence is usually used to carry out the radius-to-frequency mapping, comparing the position angle shifts on various frequencies~\citep{blaskiewicz91, hoenxil1997, mitrali2004, mitragupta2009}. The results are mostly consistent with the fact, that higher frequencies are generated closer to the stellar surface. It was also shown by~\citet{karjohnston2006} that in fact for some pulsars the $p.a.$ shift on two frequencies ($1.4$ and $3.1$ GHz) is effectively the same, hence implying the weak dependence of the $p.a.$ shift on the frequency.\par

\begin{figure}
\centering
\includegraphics[width=\columnwidth]{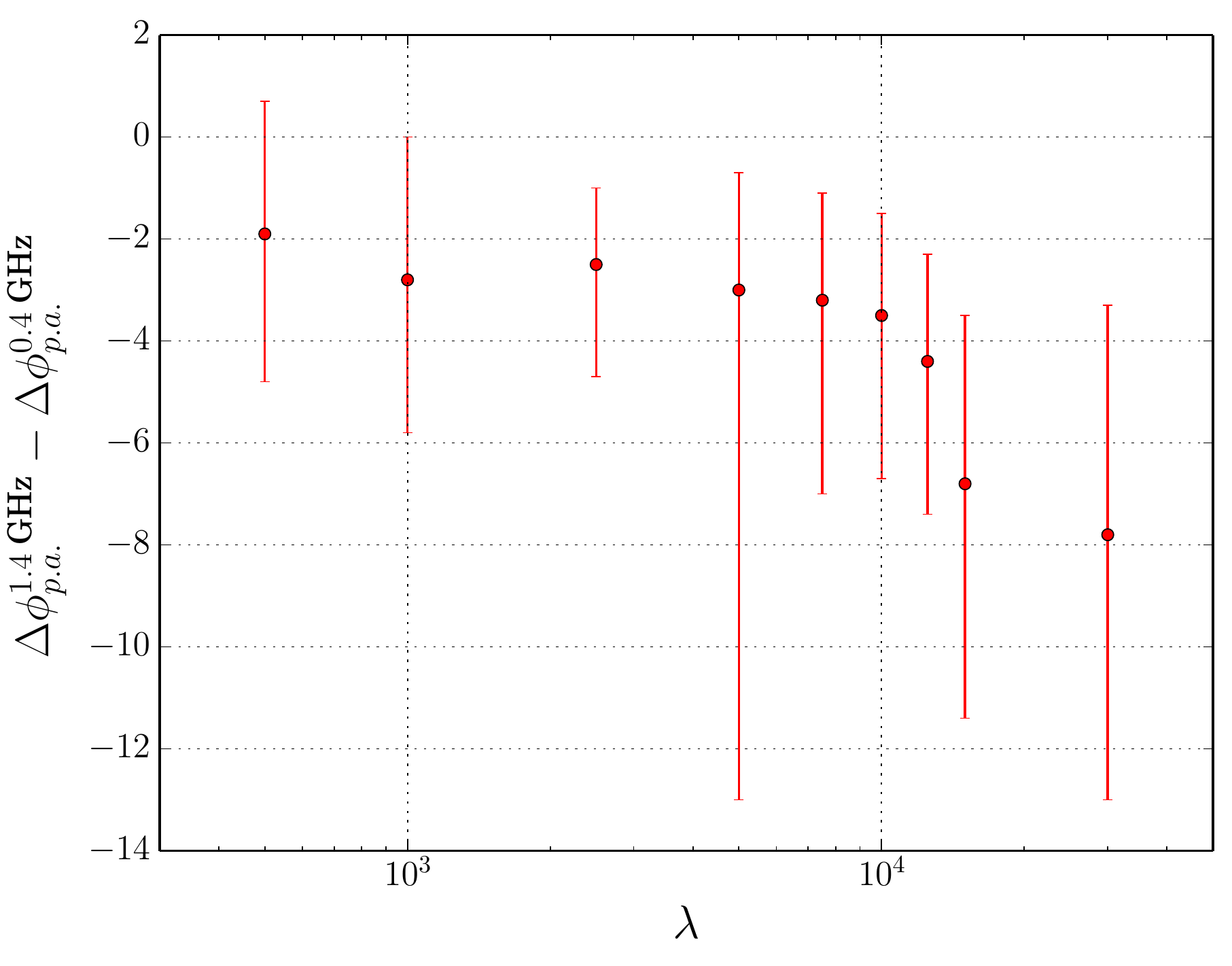}
\caption{Difference of the position angle shifts from the profile center defined on two frequencies for various multiplicity $\lambda=10^2\text{-}10^5$. This plot demonstrates, that the p.a. shifts on two frequencies are close to each other, and this fact is weakly sensitive to $\lambda$.}
\label{fig:dpalam}
\end{figure}

To study the dependence of the shift in Figure~\ref{fig:pashift} we show the position angle curves for two distinct frequencies ($0.4$ and $1.4$ GHz). Multiplicity, on the other hand, models how magnetospheric plasma affects radiation. The error bars are modeled by the radiation originating from various altitudes, as in observations we are not able to distinguish the emission heights. In Figure~\ref{fig:dpalam} we study the dependence of the position angle shifts difference at two frequencies on the plasma multiplicity.\par 

One should note two important things here. First, the higher the multiplicity, the more different are the curves on various frequencies. This provides a possible restriction for the multiplicity from multifrequency observations. On the other hand, despite the distinct frequencies, the curves are close to each other (see blue points to the right) and when taking into account the scattering due to emission height, they mostly overlap (even for large $\lambda$). This fact demonstrates, that the propagation effects do not necessarily imply a strong dependence of the $p.a.$ shift on frequency, especially if the radiation is generated in a wide range of heights. 

\section{Discussions and conclusion}

In this paper we demonstrate that complex behaviour of pulsar light-curves and polarization profiles can be explained with a propagation theory assuming a different loci in the parameter space: pulsar inclination geometry, emission region, plasma multiplicity and mean Lorentz-factor. Some general properties of the mean profile formation are also discussed. \par

At first, in Sect.~\ref{sec:vsign} we explain how the sign of the circular polarization of a single mode can change over a profile. For usual parameters of the magnetospheric plasma the polarization is being formed high enough, so the sign of the $V$ is governed by the derivative of the phase $\delta$ (appearing due to nonzero electric field in the pulsar magnetosphere) and, thus, is fixed. In some cases, however, when the altitude at which polarization becomes frozen is low, one can have a sign that depends on the rotation phase.\par

In Sect.~\ref{sec:anomal} the possible explanations for complex directivity patterns of some pulsars are discussed. While most of the pulsars follow the simple hollow cone model directivity pattern, some clearly contradict with it. We show that assuming the generation of X and O modes on various altitudes one can easily explain this behaviour. On the other hand, in Sect.~\ref{sec:chump} we have shown that there is no need to assume anomalous altitude profile of radiation for some two-peaked pulsars, that have a hump in the center of the profile (as was done by~\citealt{mitra2004}). Such effect can be easily explained by the suppression of the plasma density near the center of the directivity pattern, as in this case we will have a weaker shift from the RVM curve.\par

We further discussed the more general properties. In Sect.~\ref{sec:dirpatt} the directivity patterns for various multiplicities $\lambda$ and mean Lorentz-factors of the secondary plasma $\gamma_0$ were presented. The role of plasma cyclotron absorption in formation of the mean profiles was also discussed. It was shown, that one can obtain single-peaked pulsars for different impact angles $\beta$, and that the only reliable way to distinguish between those cases is to analyze the polarization curves. On top of that in Sect.~\ref{sec:interpulse} the pulsars with interpulses are discussed. We demonstrate the directivity patterns for various obliquity angles and show the formation of the main pulse and interpulse and their polarization curves for various impact angles. \par

Further, in Sect.~\ref{sec:radtofreq} we discuss the possible explanation of the position angle curve width for two characteristic pulsars, showing the possibility to determine the altitudes and sizes of the emitting region. It is shown that the altitudes are in a good agreement with the results obtained within the simple geometric and A/R effects considered by~\citet{blaskiewicz91} and \citet{mitragupta2009}. But upon that, our method provides an additional information about the width of the radiating region.\par

Finally, we analyze the frequency dependence of the shift of the position angle curve from the center of the mean profile. One of the key arguments against the importance of the propagation effects in the magnetosphere is that for some pulsars the position angle curve does not strongly depend on the observation frequency~\citep{karjohnston2006}. In Sect.~\ref{sec:posangshift} we analyze the position angle curves for various plasma multiplicity factors on distinct frequencies (with error bars due to generation in a wide shell of altitudes). We demonstrate that in fact even a high multiplicity does not necessarily imply a strong dependence of position angle curve on frequency, and the curves for two frequencies mostly coincide.\par

In Paper III we will confront the predictions of our model with observational data. We assume that further development of the self-consistent technique discussed above will allow us to make a powerful tool to estimate the plasma parameters for individual pulsars. \par

\section*{Acknowledgments}
The observational data for PSR~J0738-4042 (at 1375 MHz) and PSR~J1022+1001 (at 728 MHz) were obtained from the EPN database under the Creative Commons Attribution 4.0 International licence. The numerical code that supports the plots and diagrams within this paper as well as the relevant parameters of modeling are available from the authors upon reasonable request.\par 

We thank Ya.N.~Istomin, B.~Stappers and P.~Weltevrede for their interest and useful discussions and the anonymous referee for instructive comments which helped us to improve the manuscript. This work was partially supported by Russian Foundation for Basic Research (Grant no. 14-02-00831). AAP is supported by Porter Ogden Jacobus Fellowhip, awarded by the graduate school of Princeton University.

\bibliographystyle{mnras}
\bibliography{references}


\bsp

\appendix

\section{Modified plasma density profile}
\label{sec:density}

\begin{figure}
\centering
\includegraphics[width=\columnwidth]{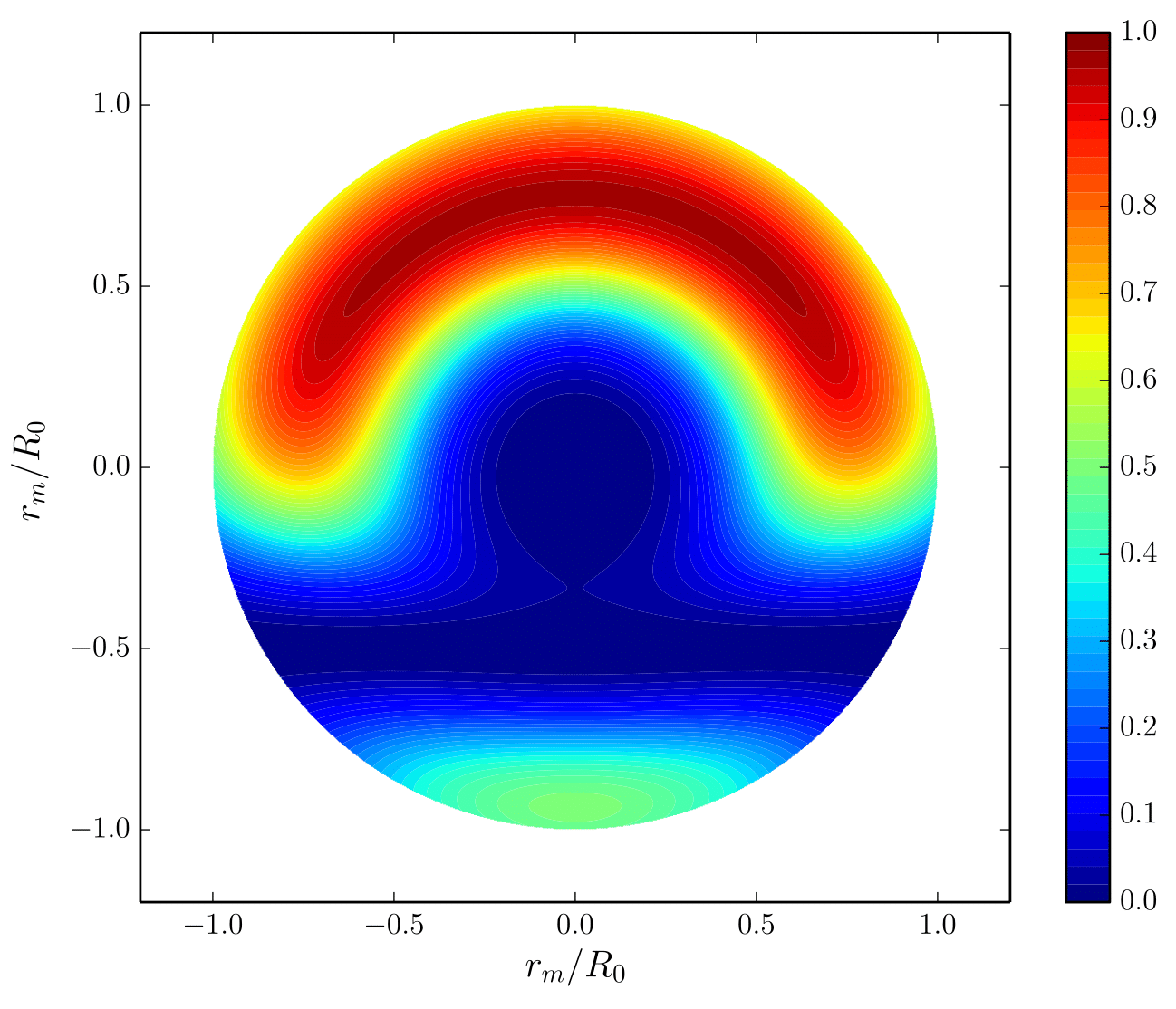}
\caption{Density profile for inclination angles $\alpha\approx90\degr$. Particle creation is suppressed both near the rotation axis (where the curvature of magnetic field lines is too large) and near the $\bmath{\Omega}\cdot\bmath{B}=0$ line (where the potential drop through the gap is too low).}
\label{fig:dens}
\end{figure}

In Paper I the axisymmetric distribution of outflowing plasma within polar cap was assumed. In general this assumption is incorrect and this fact may be important for almost orthogonal radio pulsars, when Goldreich-Julian charge density  $\rho_{\rm GJ} = -\bmath{\Omega} \cdot \bmath{B}/2\pi c$ changes the sign within the polar cap. Indeed, near this line the potential drop through the gap (which is proportional to Goldreich-Julian charge density $\rho_{\rm GJ}$) is too low to create pairs. 

For this reason axisymmetric density profile considered in Paper I is adjusted by the empiric Gaussian factor, that depends on polar angle $\theta$ from the rotation axis $\bmath{\Omega}$. As a result, within the polar cap in the vicinity of the neutron stellar surface we obtain (see Figure~\ref{fig:dens})
\begin{equation}
\label{eq:profile}
    g(r_{m},\varphi_{m}) = \frac{\exp{(-r_m^4/R_{0}^{4})}}{1+\left(r_0/r_m\right)^{5}}\left(1-\exp{\left[-\frac{\left(\pi/2-\alpha + \theta_{m}\right)^2}{2(\delta\theta)^2}\right]}\right).
\end{equation}
Here $r_{m} < R_{0}$ and $\varphi_{m}$ are the polar coordinates, $R_{0} = (\Omega R/c)^{1/2}R$ is the polar cap radius, $r_{0}$ determines the dimension of a 'hole' in the hollow cone, $\theta_{m} = 2/3 (r_{m}/R)\sin \varphi_{m}$ and $\delta\theta$ is the empirical angular width of the gap near the $\bmath{\Omega}\cdot\bmath{B}=0$ line. The first factor in (\ref{eq:profile}) models the suppression of secondary plasma generation near the magnetic axes $r_{m}\lesssim r_0$ where the magnetic field lines have large curvature radius, while the second one corresponds to zero line $\bmath{\Omega}\cdot\bmath{B}=0$.
It is important that the line $\bmath{\Omega}\cdot\bmath{B}=0$ at the star surface
locates below the magnetic pole. This implies that the appropriate region of the 
dirrectivity pattern (which is formed at the distances $r \gg R$) can be below 
the equator $\theta = \pi/2$. As a result, the interpulse can be connected this the similar radiation domain as the main one. 

\section{Magnetic field structure}
\label{sec:anticurrent}

As it was demonstrated in Paper I, the structure of the pulsar magnetosphere obtained numerically by many authors at first within force-free approximation~\citep{spitkovsky2006, kalcont2009, petri2012} and later within MHD~\citep{tchekhovskoy2013} and even PIC simulations~\citep{philippovspitkovsky2014} can be modelled good enough by rotating dipole magnetic field and radial quasi-monopole analytical solutions obtained by~\citet{michel73} and~\citet{bogovalov99}. The transition between these two asymptotic behaviors takes place in the vicinity of the light cylinder $R_{\rm L}$. It is this magnetic field structure that was used in our previous simulations.   

As a result, for ordinary pulsars ($P \sim 1$  s) for which Eqn. (\ref{eq:resc}) gives $r_{\rm esc} \ll R_{\rm L}$ the polarization characteristics are determined by the domain with almost dipole magnetic field. In this case the mean profiles are well described by 'hollow cone' model with the S-shape curve of the $p.a.$ dependence on the phase $\phi$. On the other hand, according to (\ref{eq:resc}), especially for millisecond pulsars ($r_{\rm esc} \gg R_{\rm L}$) the dipole magnetic field in the polarization formation domain should be adjusted to quasi-radial (and, hence, homogeneous) wind component. As will be shown in Paper III, more than a half of millisecond pulsars have an approximately constant $p.a.$ within the main pulse.  

On the other hand, as was recently obtained by~\citet{tchekhphil2016}, for large enough inclination angles $\alpha > 30\degr$ the angular structure of the radial wind differs drastically from the Michel-Bogovalov 'split-monopole' solution. For this reason below we use the following expressions for the magnetic field in the wind domain
\begin{equation}
\begin{aligned}
B_{r} & = \frac{\Psi_{\rm tot}}{2 \pi r^2}
\left(\cos^2\alpha + \pi \sin\theta \cos\varphi \sin^2\alpha \right), \\
B_{\varphi} & = -\frac{\Psi_{\rm tot}}{2 \pi r R_{\rm L}}\left(\sin\theta \cos^2\alpha + \pi \sin^2\theta \cos\varphi \sin^2\alpha \right),
\end{aligned}
\end{equation}
where $\Psi_{\rm tot} = \pi f_{*} R^{2}(\Omega R/c) B_{0}$ is the total magnetic flux in the wind and $1.592 < f_{*}(\alpha) < 1.96$ is the dimensionless polar cap area~\citep{bgi83, tchekhphil2016}. Here the first terms correspond to  analytical ''split monopole'' solution and the second ones correspond to orthogonal magnetic structure obtained numerically by~\citet{tchekhphil2016}.

\label{lastpage}

\end{document}